\documentclass[twocolumn,usenatbib]{aastex63}
\usepackage{lipsum}
\usepackage{totcount}
\usepackage{comment}
\usepackage{amsmath}
\newtotcounter{citnum} 
\def\oldbibitem{} \let\oldbibitem=\bibitem
\def\bibitem{\stepcounter{citnum}\oldbibitem}

\usepackage{floatrow}
\usepackage[caption=false]{subfig}
\newcommand{\degree}{\mbox{$^{\circ}$}}

\graphicspath{{./}{}}

\submitjournal{AJ}

\shorttitle{Decaying Orbit of the hot Jupiter WASP-12b}
\shortauthors{J.D. Turner, A. Ridden-Harper, R. Jayawardhana}

\newcommand\NewPeriod{1.091420090$\pm$0.000000041 days}
\newcommand\Newdecayrate{32.53$\pm$1.62 msec year$^{-1}$}

\begin{document}

\title{Decaying Orbit of the Hot Jupiter WASP-12b: Confirmation with TESS Observations}

\correspondingauthor{Jake D. Turner}
\email{astrojaketurner@gmail.com, jaketurner@cornell.edu}

\author[0000-0001-7836-1787]{Jake D. Turner}
\affil{Department of Astronomy and Carl Sagan Institute, Cornell University, Ithaca, New York 14853, USA}

\author[0000-0002-5425-2655]{Andrew Ridden-Harper}
\affil{Department of Astronomy and Carl Sagan Institute, Cornell University, Ithaca, New York 14853, USA}

\author[0000-0001-5349-6853]{Ray Jayawardhana}
\affil{Department of Astronomy, Cornell University, Ithaca, New York 14853, USA}



\begin{abstract}
Theory suggests that the orbits of some close-in giant planets should decay due to tidal interactions with their host stars. To date, WASP-12b is the only hot Jupiter reported to have a decaying orbit, at a rate of 29$\pm$2 msec year$^{-1}$. We analyzed data from NASA's Transiting Exoplanet Survey Satellite (TESS) to verify that WASP-12b's orbit is indeed changing. We find that the TESS transit and occultation data are consistent with a decaying orbit with an updated period of \NewPeriod{} and a decay rate of \Newdecayrate{}. 
We find an orbital decay timescale of $\tau$ = P/$|\dot P|$ = 2.90$\pm$0.14 Myr. If the observed decay results from tidal dissipation, the modified tidal quality factor is Q'$_{\star}$ = 1.39$\pm$0.15 $\times 10^{5}$, which falls at the lower end of values derived for binary star systems and hot Jupiters. Our result highlights the power of space-based photometry for investigating the orbital evolution of short-period exoplanets. 

\end{abstract}

\keywords{planets and satellites: gaseous planets --  planet-star interactions }


\section{Introduction} \label{sec:intro}

Some exoplanetary systems exhibit temporal variations in their orbital parameters. Such systems provide valuable insights into their physical properties and the processes that cause the variations. Light curves of transiting planets can be used to search for transit timing variations (TTVs; \citealt{Agol2005,2018haexAgol}) and, less frequently, for variations in transit duration (\citealt{2018haexAgol}) and the impact parameter (\citealt{Herman2018,Szabo2020MNRAS}). The presence of TTVs can indicate additional bodies in the system or an unstable orbit resulting from tidal forces of the star (e.g., \citealt{Miralda2002}; \citealt{Mazeh2013}). Also, TTVs in tightly packed multi-planetary systems have been detected and used to constrain the masses of planets in the system (e.g., \citealt{2018haexAgol}).

Theory suggests that short-period massive planets orbiting stars with surface convective zones may exchange energy with their host stars through tidal interactions, causing the host star to spin faster and the planet’s orbit to decay (e.g., \citealt{Lin1996,Chambers2009,Lai2012,Penev2014,Barker2020}). As the planet’s orbit decays over millions of years, its orbital period will change by a small, yet potentially detectable amount. Observations of this decay are now possible because some hot Jupiter systems have been monitored for decades. Such measurements enhance our understanding of the hot Jupiter population (e.g. \citealt{Jackson2008,Hamer2019}).

\begin{figure*}[htb]
\includegraphics[width=0.80\textwidth]{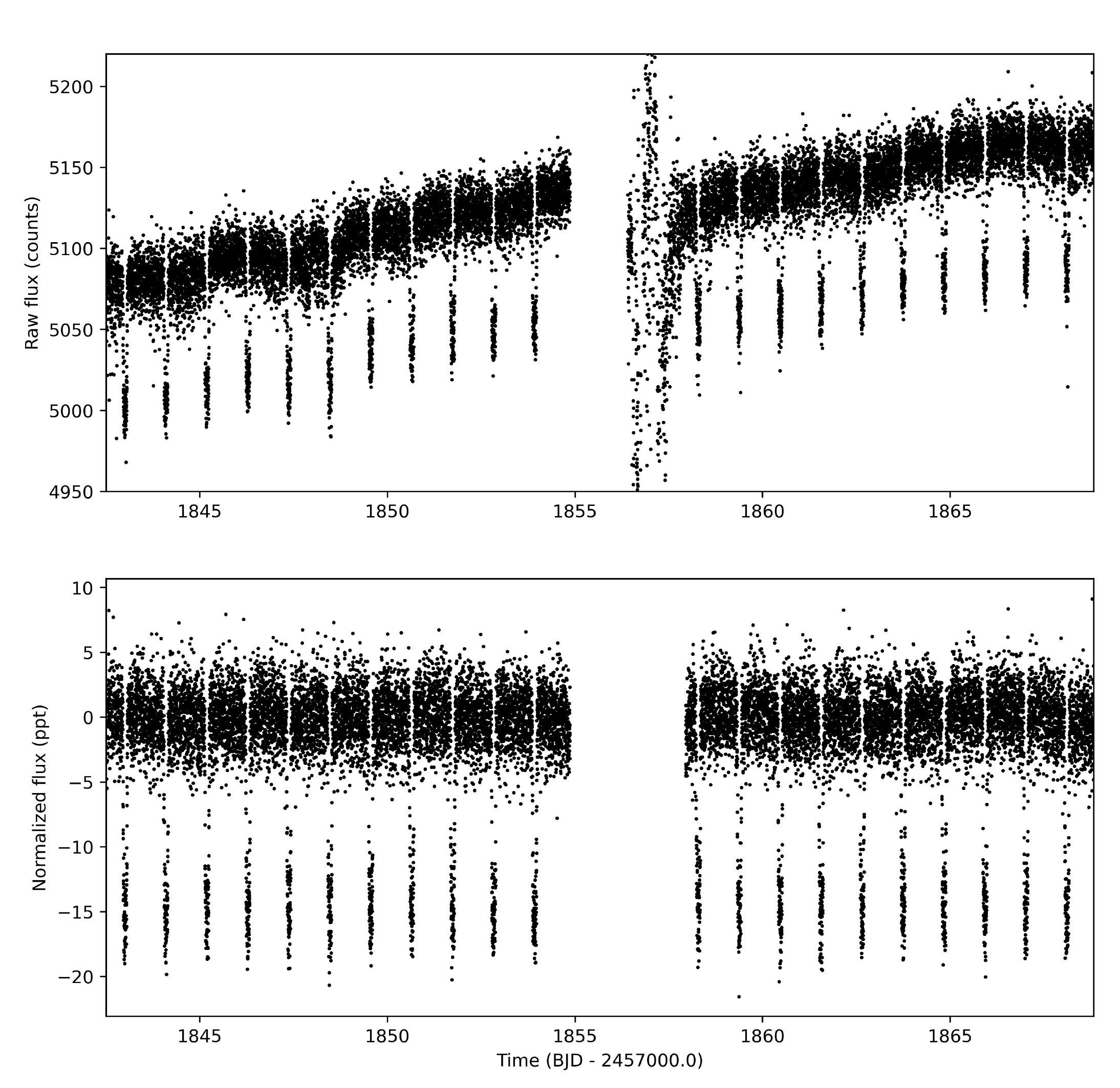}
\caption{TESS light curve of WASP-12b in Sector 20. Top: Raw simple aperture photometry light curves. Bottom: Detrended Presearch Data Conditioning (PDC) time series. }
\label{fig:lightcurves}
\end{figure*}

Currently, WASP-12b is one of the few hot Jupiters confirmed to have a varying period. It orbits a G0 star with a short period of one day (\citealt{Hebb2009}). Since its discovery in 2009, WASP-12b's orbital parameters (e.g., \citealt{Campo2011,Maciejewski2016}) and atmosphere (e.g., \citealt{Croll2011,Stevenson2014,Sing2016}) have been studied extensively. Observations with the Cosmic  Origins  Spectrograph  (COS) on the Hubble Space Telescope reveal an early ingress in the near-ultraviolet (\citealt{Fossati2010a,Haswell2012,Nichols2015}). An escaping atmosphere was suggested as the cause of the early ingress (e.g., \citealt{Lai2010,Bisikalo2013B,Turner2016a}). \citet{Maciejewski2016} were the first to report evidence of a decreasing orbital period for WASP-12b. At the time, the cause of the changing orbit was uncertain, and could be ascribed to orbital decay, apsidal precession, or the Romer effect. Follow-up observations confirmed the changing orbital period, with models slightly favoring orbital decay as the cause  (\citealt{Patra2017,Maciejewski2018,Bailey2019}). Recently, using new transit, occultation and radial velocity observations, \citet{Yee2020} presented strong evidence that WASP-12b's period variation is caused by orbital decay. The decay rate has been constrained to between 20-30 milliseconds/yr. 

Motivated by these indications of WASP-12b's changing period, we use NASA's Transiting Exoplanet Survey Satellite (TESS; \citealt{Ricker2015}) observations to verify its orbital decay, and derive updated orbital parameters and planetary properties. TESS is well suited for our study because it provides high-precision time-series data, ideal for searching for TTVs (e.g. \citealt{Hadden2019}; \citealt{vonEssen2020}; \citealt{RiddenHarper2020}).

\section{Observations and Data Reduction}
TESS observed WASP-12b in Sector 20 (December 24, 2019 to January 20, 2020; Figure \ref{fig:lightcurves}). These observations were processed by the Science Processing Operations Center (SPOC) pipeline \citep{Jenkins2016}. The final product of the SPOC pipeline are light curves corrected for systematics that can be used to characterize transiting planets. All of the SPOC data products are publicly accessible from the Mikulski Archive for Space Telescopes (MAST)\footnote{\url{https://archive.stsci.edu/}}. The Presearch Data Conditioning (PDC) component of the SPOC pipeline corrects the light curves for pointing or focus related instrumental signatures, discontinuities resulting from radiation events in the CCD detectors, outliers, and flux contamination. We use the PDC light curve for the analysis in this paper, however a known issue\footnote{The issue is related to inaccurate uncertainties in the 2D black model, which represents the fixed pattern that is visible in the black level for a sum of
many exposures. See TESS Data Release Notes: Sector 27, DR38.} results in the SPOC pipeline overestimating the uncertainties for some light curves, including that of WASP-12b. Therefore, it is recommended that the uncertainties be estimated from the scatter (Barclay, T., private communication), until the data is reprocessed with an updated pipeline (data released after DR 38 should be unaffected).  To do this, we used the standard deviation of the entire out-of-transit baseline as the uncertainty. Using smaller out-of-transit baselines surrounding only a few transits did not significantly change the standard deviation, indicating that the noise is relatively constant.  The raw light curve from the SPOC and detrended PDC light curve are shown in Figure \ref{fig:lightcurves}.

\begin{figure*}
  \begin{flushleft} \textbf{(a)} \end{flushleft}\vspace{\baselineskip}
\includegraphics[width=0.7\textwidth]{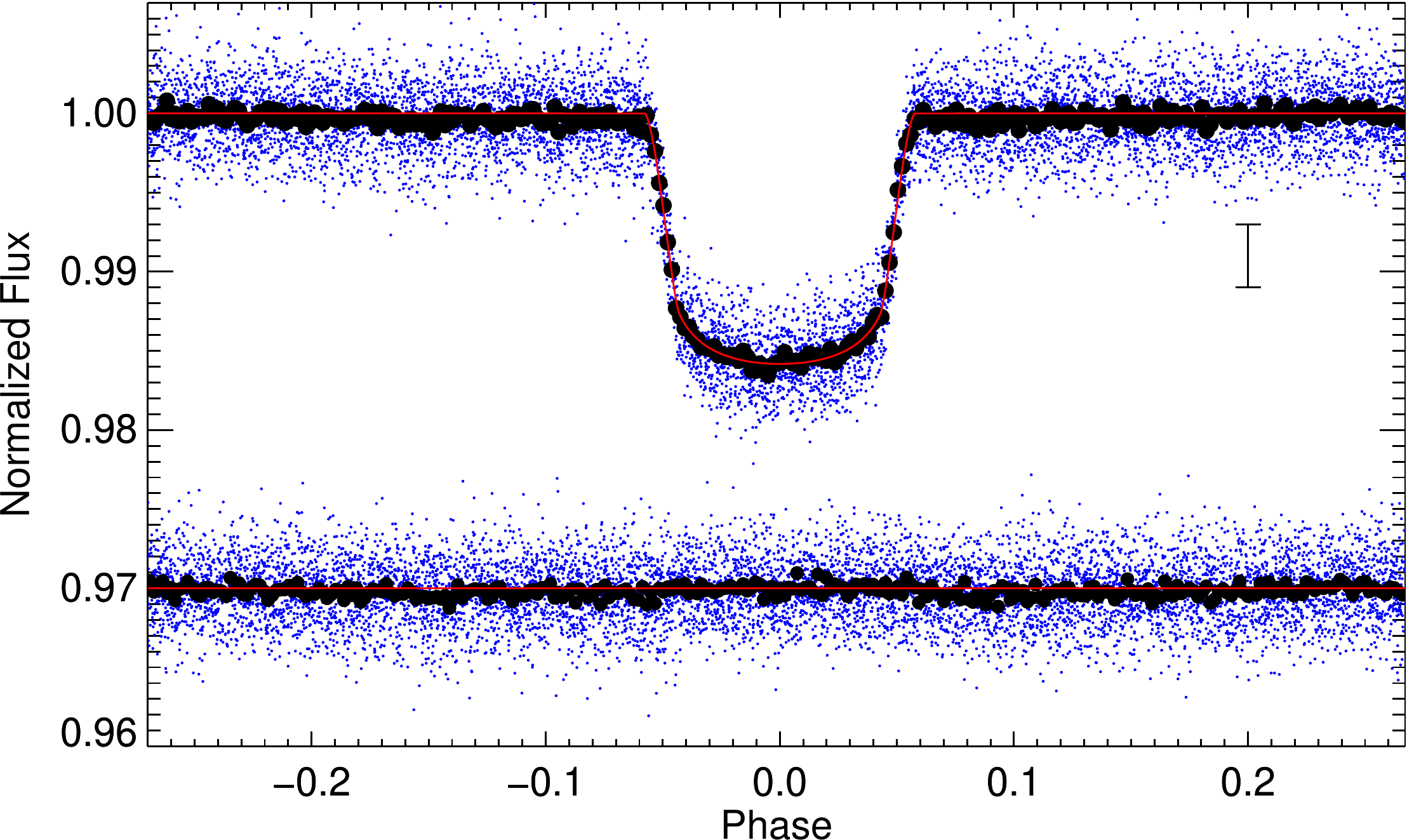}\\ 
  \begin{flushleft} \textbf{(b)} \end{flushleft}\vspace{\baselineskip}
\includegraphics[width=0.7\textwidth]{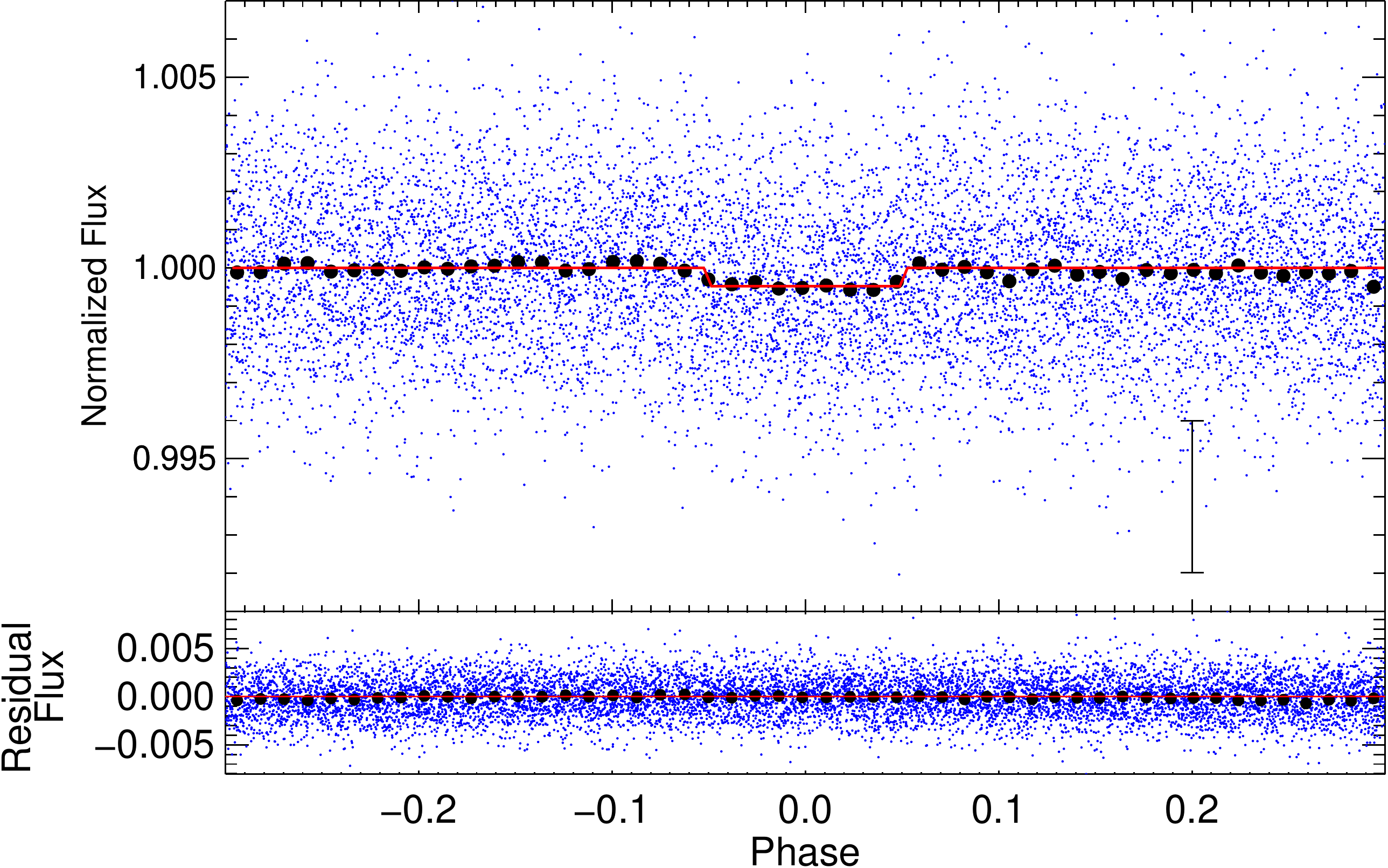}\\ 
\caption{Phased folded transit (panel a) and occulation (panel b) light curves of WASP-12b from TESS. The unbinnned and binned data are shown in blue and black, respectively. The binned transit and occulation data have been binned by 2 and 18 minutes, respectively. The best-fitting model obtained from the EXOplanet MOdeling Package (\texttt{EXOMOP}) is shown as a solid red line. The residuals (light curve - model) are shown below the light curve.} 
\label{fig:modelfit}
\end{figure*}

\section{Data Analysis}
\subsection{Transit Modeling}
We modeled the TESS transits of WASP-12b with the EXOplanet MOdeling Package (\texttt{EXOMOP}; \citealt{Pearson2014,Turner2016b,Turner2017})\footnote{\texttt{EXOMOPv7.0}; \href{https://github.com/astrojake/EXOMOP}{https://github.com/astrojake/EXOMOP} } to find a best-fit. \texttt{EXOMOP} creates a model transit using the analytic equations of \cite{Mandel2002} and the data are modeled using a Differential Evolution Markov Chain Monte Carlo (DE-MCMC; \citealt{Eastman2013}) analysis. To account for red noise in the light curve, \texttt{EXOMOP} incorporates the residual permutation, time-averaging, and wavelet methods. More detailed descriptions of \texttt{EXOMOP} can be found in \citet{Pearson2014} and \citet{Turner2016}.

Each transit in the TESS data (lower panel of Figure \ref{fig:lightcurves}) was modeled with \texttt{EXOMOP} independently. We used 20 chains and 20$^{6}$ links for the DE-MCMC model and ensure chain convergence (\citealt{Ford2006}) using the Gelman-Rubin statistic (\citealt{Gelman1992}). The mid-transit time ($T_{c}$), planet-to-star radius ($R_{p }/R_{*}$), inclination ($i$), and scaled semi-major axis ($a/R_{*}$) are set as free parameters for every transit. The period ($P_{orb}$) and linear and quadratic limb darkening coefficients are fixed during the analysis. The linear and quadratic limb darkening coefficients used in the modeling were set to 0.2131 and 0.3212 (taken from \citealt{Claret2017}), respectively. 

The parameters derived for every TESS transit event can be found in Table \ref{tb:lighcurve_model_TESS}. The modeled light curves for each individual transit can be found in Figures \ref{fig:ind_transits_sec20_1}--\ref{fig:ind_transits_sec20_4} in Appendix \ref{app:individual_transits}. All parameters for each transit event are consistent within 2$\sigma$ of every other transit. 

To obtain the final fitted parameters, the light curve of WASP-12b was phase-folded at each derived mid-transit time and modeled with \texttt{EXOMOP}. The phase-folded light curve and model fit can be found in Figure \ref{fig:modelfit}. We use the light curve model results combined with literature values to calculate the planetary mass (M$_{b}$;  \citealt{Winn2010b}), radius (R$_{b}$), density ($\rho_{b}$), equilibrium temperature (T$_{eq}$), surface gravity ($\log{g_{b}}$; \citealt{Southworth2007a}), orbital distance ($a$), inclination ($i$), Safronov number ($\Theta$; \citealt{Safronov1972}; \citealt{Southworth2010a}), and stellar density ($\rho_{\ast a}$; \citealt{Seager2003}). The planet properties and transit ephemeris we derived for WASP-12b are shown in Table \ref{tb:planet_parameters}. All the planetary parameters are consistent with their discovery values \citep{Hebb2009} but their precision is greatly improved.      

\begin{table}
\caption{Physical  properties  of  WASP-12b derived  from  the light curve modeling of the TESS data}
    \centering
    \begin{tabular}{lcccc}
     \hline
     Parameter              &  units    & value  & 1 $\sigma$ uncertainty  \\
     \hline
  R$_p$/R$_\ast$            &               &   0.1166445    & 0.0000097\\
  a/R$_\ast$                &               &   3.1089    & 0.0014       \\
  $\delta$$_{occ}$        &              & 0.00048             & 0.00010  \\
  Inclination               & \degree       &   84.955 & 0.037  \\
  Duration         & mins          & 178.60  & 0.14 \\
 b                        &         & 0.2734             &   0.0020\\
  R$_{b}$                   & R$_{Jup}$     &  1.884 & 0.057      \\
  M$_{b}$                   & M$_{Jup}$     &  1.46  & 0.27                  \\
  $\rho_{b}$                & g cm$^{-3}$  &0.271 & 0.056                     \\
  $\log{g_{b}}$   	        &   cgs         & 2.38 & 0.39                  \\
  $\rho_{\ast a}$           & g cm$^{-3}$   & 0.477 & 0.026                                  \\
  T$_{eq}$                  &   K           & 2551 & 56                      \\
  $\Theta$                  &               &  0.0260 & 0.0053                    \\ 
  a                         &  au           &   0.02399 & 0.00072                      \\
   \hline
    \end{tabular}
    \label{tb:planet_parameters}
\end{table}

\subsection{Occultation Modeling}
The TESS occultation was also modeled with \texttt{EXOMOP} to find the best-fit parameters. The light curve we fit was obtained by phase-folding all the data about the secondary eclipse using the first TESS transit as the reference transit time. We were unable to see or fit individual occulations as they were too shallow. Again, we used 20 chains and 20$^{6}$ links for the DE-MCMC model and ensure chain convergence using the Gelman-Rubin statistic. The mid-occulation time (T$_{occ}$) and occultation depth ($\delta$$_{occ}$) were set as free parameters. The inclination, period, and scaled semi-major axis were fixed to the values obtained from the phase-folded transit curve (Table \ref{tb:planet_parameters}). The linear and quadratic limb darkening coefficients were both set to zero. We find a T$_{occ}$ = 2458843.55034$\pm$0.00088 BJD$_{TDB}$ for the first epoch (E=2325) in the TESS observations and $\delta$$_{occ}$=0.00048$\pm$0.00010. The results of the analysis do not change whether we use the first epoch or the middle epoch of the TESS data.  

\subsection{Timing}

For the timing analysis, we combined the TESS transit and occultation data with the all the prior transit and occultation times complied by \citet{Yee2020}. All the transit and occultation times used in this analysis can be found in Table \ref{tab:all_times}. This table is available in its entirety in machine-readable form online.

Similar to what was done in \citet{Yee2020} and \citet{Patra2017}, we fit the timing data to three different models. The first model is the standard constant period formalization:
\begin{align}
    t_{\text{tra}}(E) =& T_{c,0} + P_{\text{orb}} \times E,  \\
   t_{\text{occ}}(E)  =&    T_{c,0} +\frac{P_{\text{orb}}}{2}  + P_{orb} \times E,
\end{align}
\noindent where $T_{0}$ is the reference transit time, $P_{orb}$ is the orbital period, $E$ is the transit epoch, and $T_{tra}(E)$ is the calculated transit time at epoch $E$. 

The second model assumes that the orbital period is changing uniformly over time:
\begin{align}
    t_{\text{tra}}(E) =& T_{c,0} + P_{orb} \times E  + \frac{1}{2} \frac{\mathrm{d}P_{orb}}{\mathrm{d}E} E^2, \\
    t_{\text{occ}}(E) =&  T_{c,0} +\frac{P_{orb}}{2}  + P_{orb} \times E + \frac{1}{2} \frac{\mathrm{d}P_{orb}}{\mathrm{d}E} E^2,
\end{align}
where $\mathrm{d}P_{orb}/\mathrm{d}E$ is the decay rate. 

The third model assumes the planet is precessing uniformly (\citealt{Gimenez1995}):
\begin{align}
     t_{\text{tra}}(E) =&   T_{c,0} + P_{s} \times E  - \frac{e P_{a}}{\pi} \cos{\omega(N)},\\
      t_{\text{occ}}(E) =&  T_{c,0} + \frac{P_{orb}}{2} + P_{s} \times E  + \frac{e P_{a}}{\pi} \cos{\omega(N)},\\
      \omega(N) =& \omega_{0} + \frac{\mathrm{d} \omega}{\mathrm{d}E} E,\\
      P_{s} =& P_{a} \left( 1 - \frac{1}{2\pi} \frac{\mathrm{d} \omega}{\mathrm{d}E} \right),
\end{align}
where $e$ is a nonzero eccentricity, $\omega$ is the argument of pericenter, P$_{s}$ is the sidereal period and P$_{a}$ is the anomalistic period.

\begin{table}[htb]
    \centering
    \begin{tabular}{ccccc}
    \hline
    Event  & Midtime     & Error & Epoch & Source \\
    Type   & BJD$_{TDB}$ & days  &  \\
    \hline
tra	&	2458843.00493	&	0.00054	&	2325	&	This Paper	\\
tra	&	2458844.09660	&	0.00052	&	2326	&	This Paper	\\
tra	&	2458845.18785	&	0.00046	&	2327	&	This Paper	\\
tra	&	2458846.27971	&	0.00054	&	2328	&	This Paper	\\
tra	&	2458847.37083	&	0.00051	&	2329	&	This Paper	\\
tra	&	2458848.46238	&	0.00049	&	2330	&	This Paper	\\
tra	&	2458849.55308	&	0.00043	&	2331	&	This Paper	\\
tra	&	2458850.64512	&	0.00053	&	2332	&	This Paper	\\
tra	&	2458851.73590	&	0.00057	&	2333	&	This Paper	\\
tra	&	2458852.82819	&	0.00038	&	2334	&	This Paper	\\
tra	&	2458853.91924	&	0.00047	&	2335	&	This Paper	\\
tra	&	2458858.28450	&	0.00061	&	2339	&	This Paper	\\
tra	&	2458859.37732	&	0.00055	&	2340	&	This Paper	\\
tra	&	2458860.46816	&	0.00044	&	2341	&	This Paper	\\
tra	&	2458861.55803	&	0.00049	&	2342	&	This Paper	\\
tra	&	2458862.65047	&	0.00057	&	2343	&	This Paper	\\
tra	&	2458863.74203	&	0.0006	&	2344	&	This Paper	\\
tra	&	2458864.83395	&	0.00044	&	2345	&	This Paper	\\
tra	&	2458865.92421	&	0.0005	&	2346	&	This Paper	\\
tra	&	2458867.01609	&	0.00051	&	2347	&	This Paper	\\
tra	&	2458868.10763	&	0.00046	&	2348	&	This Paper	\\
occ	&	2458843.55034	&	0.00088	&	2325	&	This Paper	\\
    \hline     
    \end{tabular}
    \caption{All WASP-12b Transit and Occultation Times}
    \label{tab:all_times}
    \tablecomments{
    This table is available in its entirety in machine-readable form.
    \tablerefs{\citet{Hebb2009}, \citet{Copperwheat2013}, \citet{Croll2015}, \citet{Collins2017}, \citet{Campo2011}, \citet{Chan2011}, \citet{Cowan2012}, \citet{Crossfield2012}, \citet{Deming2015}, \citet{Fohring2013}, \citet{Hooton2019}, \citet{Kreidberg2015PASP}, \citet{Maciejewski2013,Maciejewski2016,Maciejewski2018}, \citet{Ozturk2019}, \citet{Patra2017, Patra2020}, \citet{Sada2012}, \citet{Stevenson2014}, \citet{vonEssen2019} }
    }
\end{table}

For all the three models, we found the best-fitting model parameters using a DE-MCMC analysis. We used 20 chains and 20$^{6}$ links in the model and again we ensure chain convergence using the Gelman-Rubin  statistic. The results of timing model fits can be found in Table \ref{tb:timing_models}. 

\begin{figure*}
    \centering
    \begin{minipage}{0.56\textwidth}
                          \begin{flushleft} \textbf{(a)} \end{flushleft}
                \centering
                 \subfloat{ \includegraphics[width=1\textwidth,page=1]{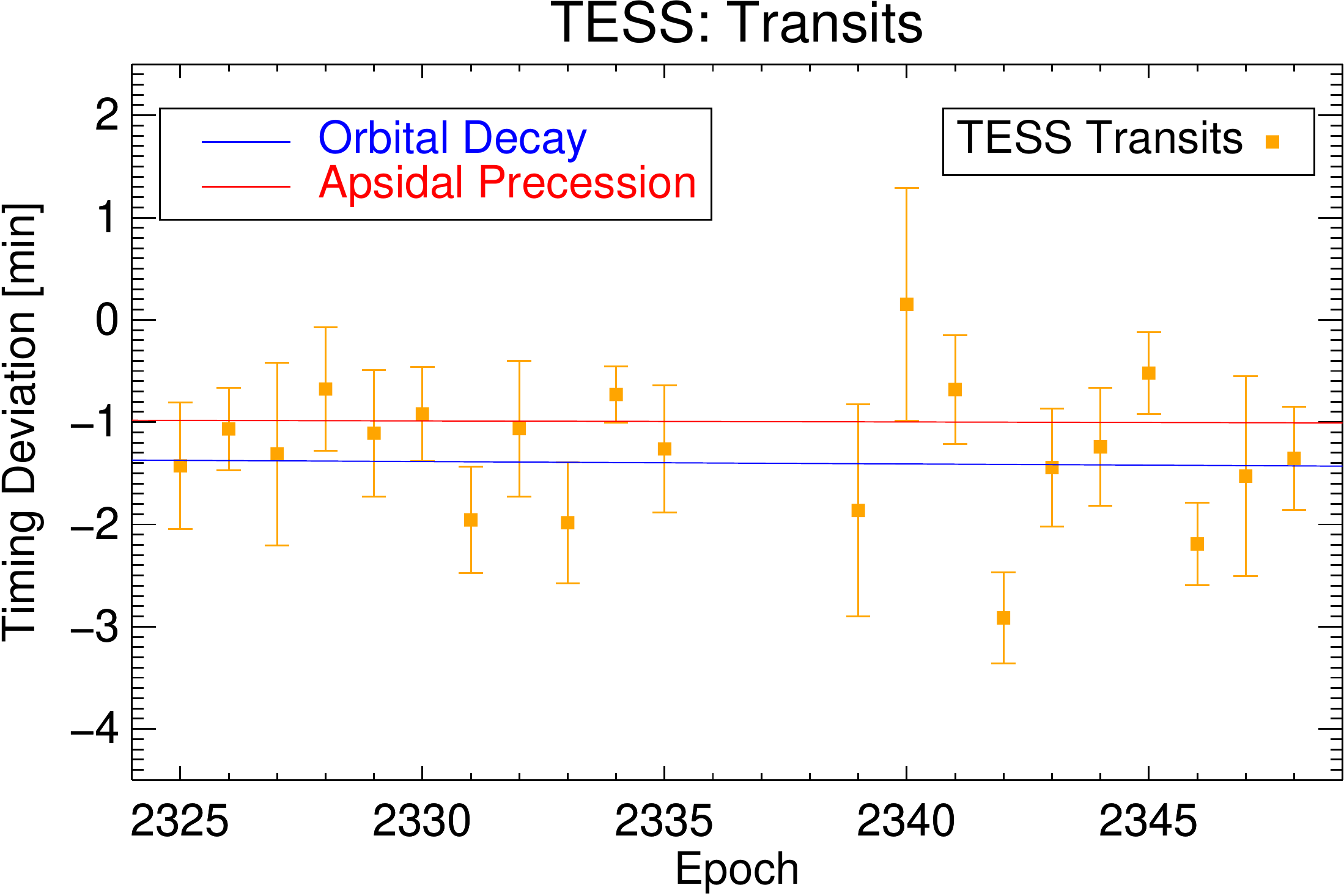}   }  \\ 
    \end{minipage}%
    \begin{minipage}{0.56\textwidth}
        \begin{flushleft} \textbf{(b)} \end{flushleft}
        \centering
              \subfloat{\includegraphics[width=1\textwidth,page=2]{TTVS_data_with_models_newV3.pdf}}\\
     \end{minipage} 
        \begin{minipage}{0.56\textwidth}
                                  \begin{flushleft} \textbf{(c)} \end{flushleft}
        \centering
              \subfloat{\includegraphics[width=1\textwidth,page=4]{TTVS_data_with_models_newV3.pdf}}\\
    \end{minipage} 
    
\caption{Transit (panels a and b) and occultation (panel c) timing variations after subtracting the data with a constant-period model. The filled black triangles are the data points complied from \citet{Yee2020} and the square orange points are from the TESS data in this paper. All the transit and occultation times can be found in Table \ref{tab:all_times}. The orbital decay and apsidal precession models are shown as the blue and red lines, respectively.
}
\label{fig:timing_fits}
\end{figure*}

Figure \ref{fig:timing_fits} shows the transit and occultation timing data fit with the orbital decay and apsidal precession models. In this figure, the best-fit constant-period model has been subtracted from the timing data. For the TESS transits alone (panel a), the orbital decay model fits the data slightly better than the precession model. When all the available transit data are combined (panel b), then it is more apparent that orbital decay is favored as the cause. The occultation data also favor the orbital decay model (panel c), with the new TESS data point important for discriminating between the two possibilities. 

\begin{table*}[!htb]
\caption{Best-Fit Parameters for Timing Models}
    \centering
    \begin{tabular}{lcccc}
     \hline
     Parameter                          &  Symbol &   units    & value  & 1 $\sigma$ uncertainty  \\
     \hline
     \textbf{Constant Period Model}\\
        Period                          & P$_{\text{orb}}$          & days         & 1.091419426 & 0.000000022   \\
        Mid-transit time                & T$_{c,0}$   &BJD$_{TDB}$     & 2456305.455519  &0.000026 \\
        \hline                                                              
        N$_{dof}$                         &            &              &182         & \\
        $\chi$$_{min}^{2}$                    &            &              & 576.08       &       \\
        BIC                             &           &               & 586.47 \\
     \hline
     \hline 
    \textbf{Orbital Decay Model}\\ 
        Period                         & P$_{\text{orb}}$        &  days         & 1.091420090 & 0.000000041 \\
        Mid-transit time               & T$_{c,0}$    &BJD$_{TDB}$      & 2456305.455795 & 0.000038 \\
        Decay Rate                      & dP/dE     &days/orbit     & -9.45$\times$10$^{-10}$ &  4.7$\times$10$^{-11}$ \\
        Decay Rate                      & dP/dt     & msec/yr       & -32.53                   & 1.62      \\
        \hline
        N$_{dof}$                         &           &               &  183    \\ 
        $\chi$$_{min}^2$                    &           &               & 188.71       \\
        BIC \                           &           &               & 204.30         \\
        \hline 
        \hline
        \textbf{Apsidal Precession Model} \\
        Sidereal Period                  & P$_{s}$    & days          &  1.091419419      &  0.000000052             \\
        Mid-transit time                 & T$_{c,0}$    & BJD$_{TDB}$     & 2456305.45481     & 0.00011                  \\
        Eccentricity                    & e         &               &  0.00363          & 0.00025 \\
        Argument of Periastron          & $\omega$$_{0}$& rad           &  2.447             &0.077               \\
        Precession Rate                 & d$\omega$/dN& rad/orbit     &  0.000963          & 0.000040                 \\
       \hline                                                          
        N$_{dof}$                         &           &               & 185                  \\ 
        $\chi$$_{min}^2$                    &           &               &212.75                  \\
        BIC \                           &           &               & 238.72                \\
        \hline 
        \hline
    \end{tabular}
    \label{tb:timing_models}
\end{table*}

Thus, we find that the orbital decay model fits the timing data the best (Table \ref{tb:timing_models}, Figure \ref{fig:timing_fits}). The fact that the constant-period model does not fit the data is consistent with previous findings (\citealt{Maciejewski2016}, \citealt{Patra2017}, \citealt{Yee2020}). The orbital decay and apsidal precession models fit the data with a minimum chi-squared ($\chi$$_{min}^{2}$) of 188.71 and 212.75, respectively. We use the Bayesian Information Criterion (BIC) to assess the preferred model. The BIC is defined as 
\begin{equation}
BIC =  \chi^{2} + k \ln{(N_{pts})}, \label{eq:BIC}
\end{equation}
where $k$ is the number of free parameters in the model fit and $N_{pts}$ is the number of data points. The power of the BIC is the penalty for a higher number of fitted model parameters, making it a robust way to compare different best-fit models. The preferred model is the one that produces the lowest BIC value. We find that the orbital decay model is the preferred model with a $\Delta$(BIC) = 34.42. We can relate the $\Delta$(BIC) and the Bayes factor $B$ assuming a Gaussian distribution for the posteriors: 
\begin{equation}
    B = \exp[-\Delta(BIC)/2] = 3\times10^{7}. 
\end{equation}
Therefore, the orbital decay model is overwhelming the preferred interpretation of the observed timing residuals. 

\section{Discussion}
From our analysis, we derive an updated period of \NewPeriod{} and a decay rate of \Newdecayrate{}. Our results indicate an orbital decay timescale of $\tau = P/|\dot P| = 2.90\pm0.14$ Myr, slightly shorter than the value derived by \cite{Yee2020} of $3.25^{+0.24}_{-0.21}$ Myr. The mass-loss rate required to explain the early ingress of WASP-12b in the near-UV is $\sim 3 \times 10^{14} g s^{-1}$, which corresponds to a mass-loss timescale of $M_p/\dot M_p \sim 300$ Myr \citep{Lai2010,Jackson2017}, which is about two orders of magnitude longer than the orbital decay timescale. Therefore, the orbital decay timescale is the dominant timescale for the evolution of WASP-12b.

Our analysis strengthens the case for the changing period of WASP-12b being produced by tidal decay. In this scenario, WASP-12b's orbital energy is dissipated by tides within its host star. The effectiveness of such tidal dissipation is quantified conveniently by the modified tidal quality factor, $Q'_{\star}$, which is related to the star's tidal quality factor, 
$Q_\star$ by 
\begin{align}
    Q'_\star = \frac{Q_\star }{\frac{2}{3} k_2},
\end{align}
where $k_2$ is a Love number. 

Assuming that the planet mass is constant, the rate of change of WASP-12b's orbital period, $\dot P$, can be related to its host star's modified tidal quality factor by the constant-phase-lag model of \cite{Goldreich1966}, defined as 
\begin{align}
    \dot P = - \frac{27\pi}{2 Q'_\star} \left ( \frac{M_p}{M_\star} \right ) \left ( \frac{R_\star}{a} \right )^5,
\end{align}
where $M_p$ is the mass of the planet, $M_\star$ is the mass of the host star, $R_\star$ is the radius of the host star and $a$ is the semi-major axis of the planet. 

By substituting our derived value of $\dot P$ and the remaining values from Table \ref{tb:planet_parameters} and \cite{Hebb2009}, we find a modified tidal quality factor of Q'$_\star$ = 1.39$\pm$0.15 $\times$ 10$^5$. This value is slightly lower than the value derived by \cite{Yee2020} of  Q'$_\star$ = 1.75$^{+0.13}_{-0.11}$ $\times$ 10$^5$, and is generally at the lower end of observed values for binary star systems \citep[$10^5 - 10^7$;][]{Meibom2015} and hot Jupiters \citep[$10^{5} -10^{6.5}$;][]{Jackson2008,Husnoo2012,Barker2020}. Additionally, \citet{Hamer2019} found that hot Jupiter host stars tend to be young, implying that such planets spiral into their stars during the main sequence phase, requiring Q'$_\star$ $\lesssim$ 10$^7$.    

It is not yet clear how to account for the observed low values of $Q'_\star$ because, in general, theoretically derived values for $Q'_\star$ tend to be higher ($10^7 - 10^{10}$; \citealt[and references therein]{Ogilvie2014}). \citet{Weinberg2017} show that if WASP-12 were a sub-giant star, nonlinear wave-breaking of the dynamical tide near the stellar core would lead to $Q'_\star \sim 2 \times 10^5$, which is reasonably close to the observed value of $Q'_\star$ = 1.56 $\times 10^5$.  However, \citet{Bailey2019} found that the observed characteristics of WASP-12 are more consistent with it being a main-sequence star, so more theoretical work could be informative. 

WASP-12b is the only exoplanet for which there is robust evidence of tidal orbital decay. However, WASP-103b, KELT-16b, and WASP-18b are also predicted to exhibit comparable rates of tidal decay \citep{Patra2020}. Hence, additional data could reveal whether they indeed exhibit hitherto undetected tidal decay or whether the theoretical predictions need to be improved. Timing observations of additional systems are warranted because they help us understand the formation, evolution and ultimate fate of hot Jupiters (e.g. \citealt{Jackson2008,Hamer2019}).

\section{Conclusions}
We analyzed TESS data of WASP-12b to characterize the system and to verify that the planet is undergoing orbital decay. Our TESS transit and occultation timing investigations confirm that the planet's orbit is changing. We compare our timing residuals to orbital decay and apsidal precession models, and our analysis highly favors the orbital decay scenario with a Bayes factor of 3$\times10^{7}$. We find an updated period of \NewPeriod{} and a decay rate of \Newdecayrate{}. Our finding indicates an orbital decay lifetime of 2.90$\pm$0.14 Myr, shorter than the estimated mass-loss timescale of 300 Myr. We also update the planetary physical parameters and greatly improve on their precision. Our study highlights the power of long-term high-precision (both in flux and timing accuracy) ground and space-based transit and occultation observations for understanding orbital evolution of close-in giant planets.

\acknowledgments

We thank Dong Lai for useful discussions. We were inspired to pursue this project after attending a TESS hackathon hosted by the Carl Sagan Institute. This paper includes data collected by the TESS mission, which are publicly available from the Mikulski Archive for Space Telescopes  (MAST). Funding for the TESS mission is provided by the NASA Explorer Program.

This research has made use of the Extrasolar Planet Encyclopaedia, NASA's Astrophysics Data System Bibliographic Services, and the the NASA Exoplanet Archive, which is operated by the California Institute of Technology, under contract with the National Aeronautics and Space Administration under the Exoplanet Exploration Program.  

We also thank the anonymous referee for their comments.


%

\facilities{\textit{TESS} (\citealt{Ricker2015}); \textit{Exoplanet Archive}}

\software{
          \texttt{EXOMOP} \citep{Pearson2014,Turner2016,Turner2017}; \texttt{IDL Astronomy Users Library} \citep{Landsman1995}; \texttt{Coyote IDL:} created by David Fanning and now maintained by Paulo Penteado (JPL)
}



\appendix

\section{Transit fits to individual TESS transit events} \label{app:individual_transits}
\begin{figure*}
\centering
\begin{tabular}{cc}
 \includegraphics[width=0.50\textwidth]{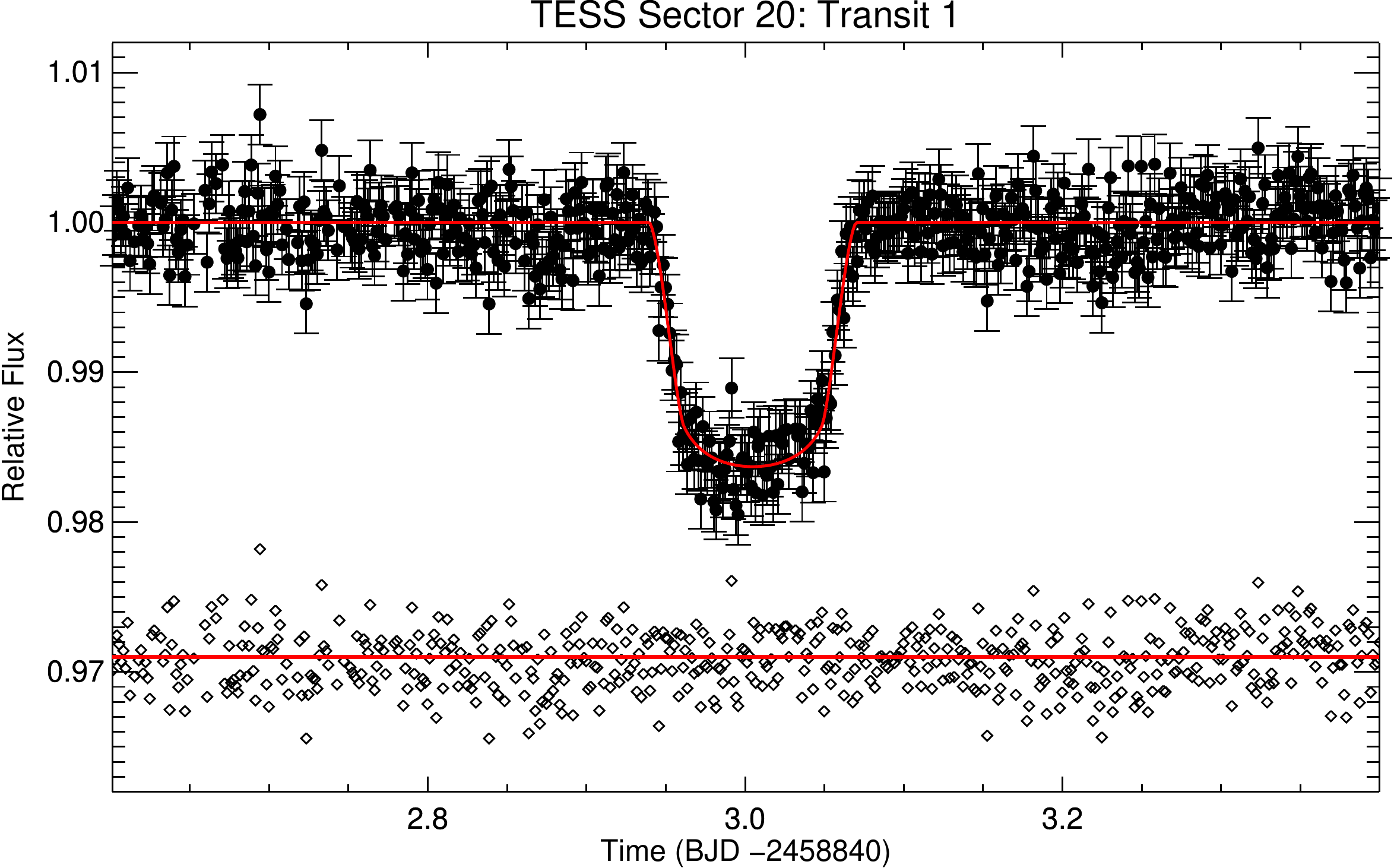} & \includegraphics[width=0.50\textwidth]{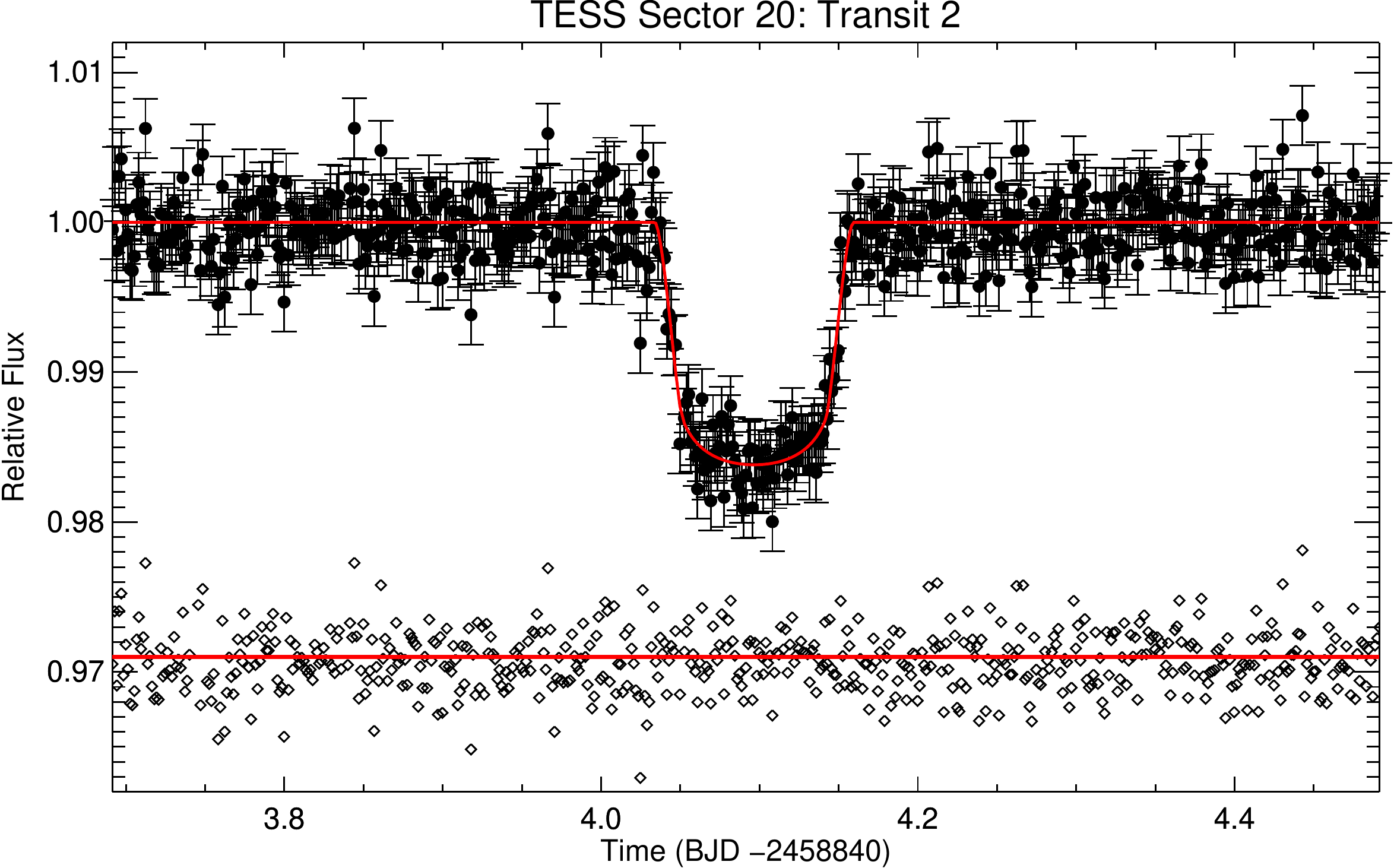}\\
 \includegraphics[width=0.50\textwidth]{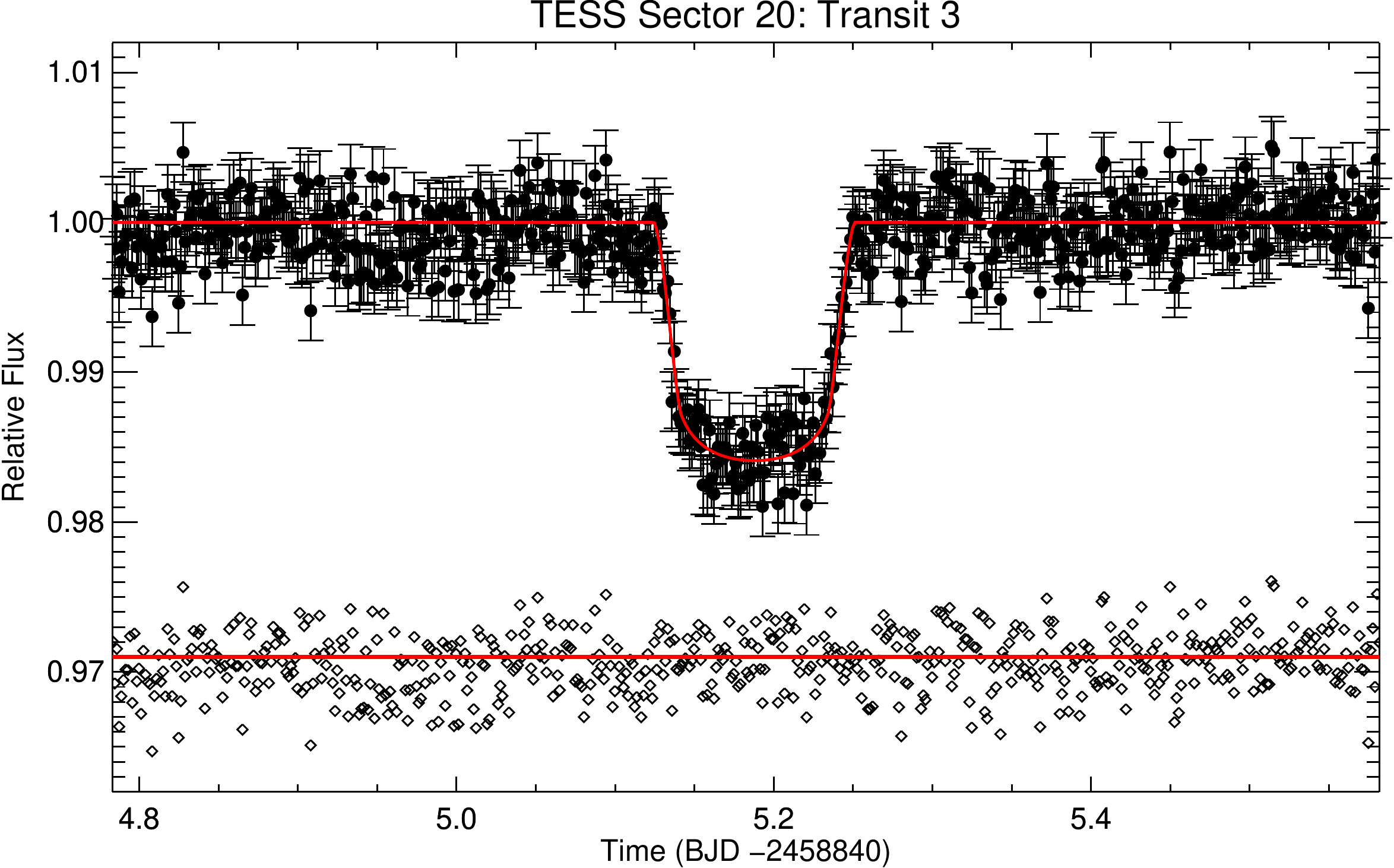} & \includegraphics[width=0.50\textwidth]{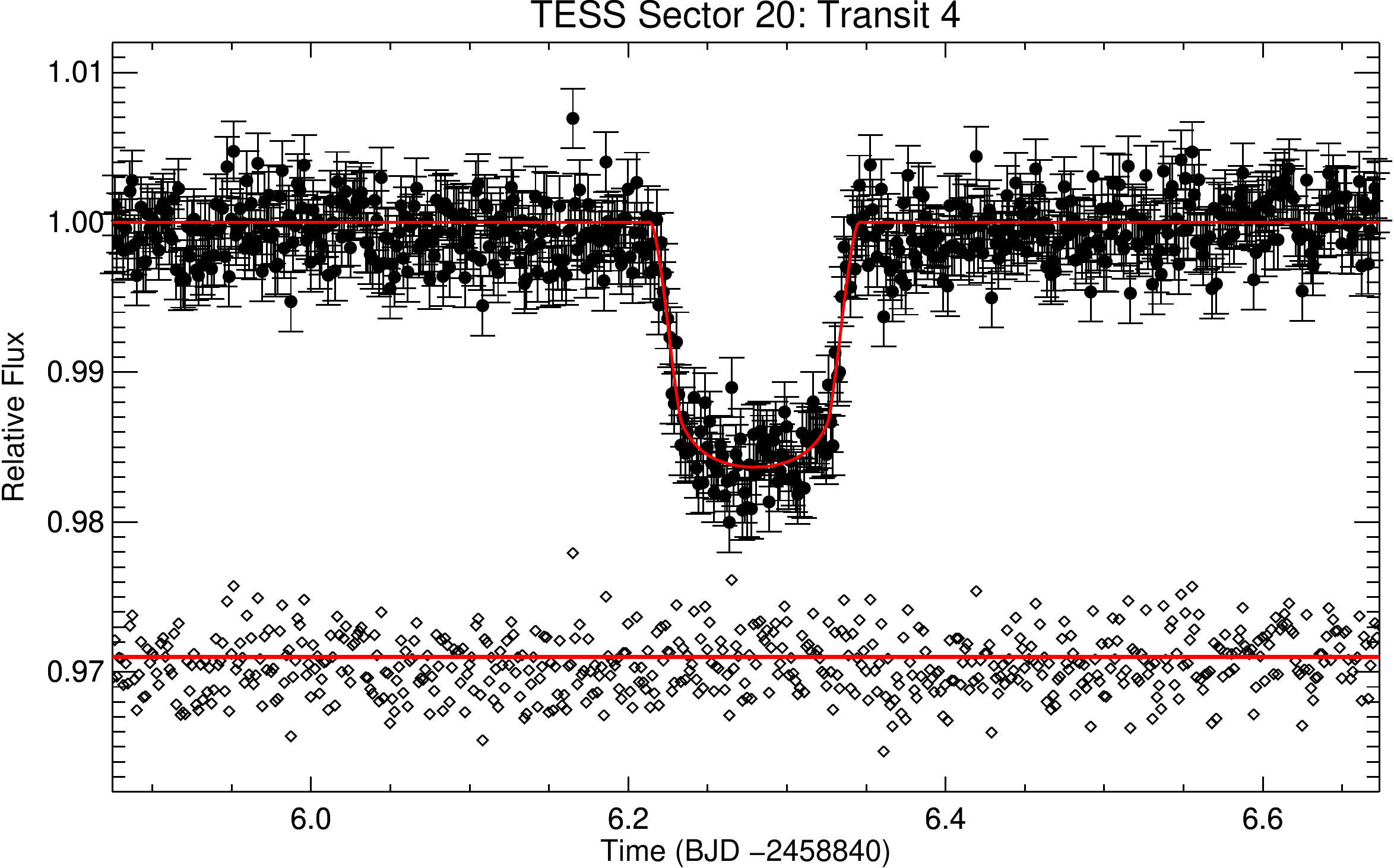}\\
  \includegraphics[width=0.50\textwidth]{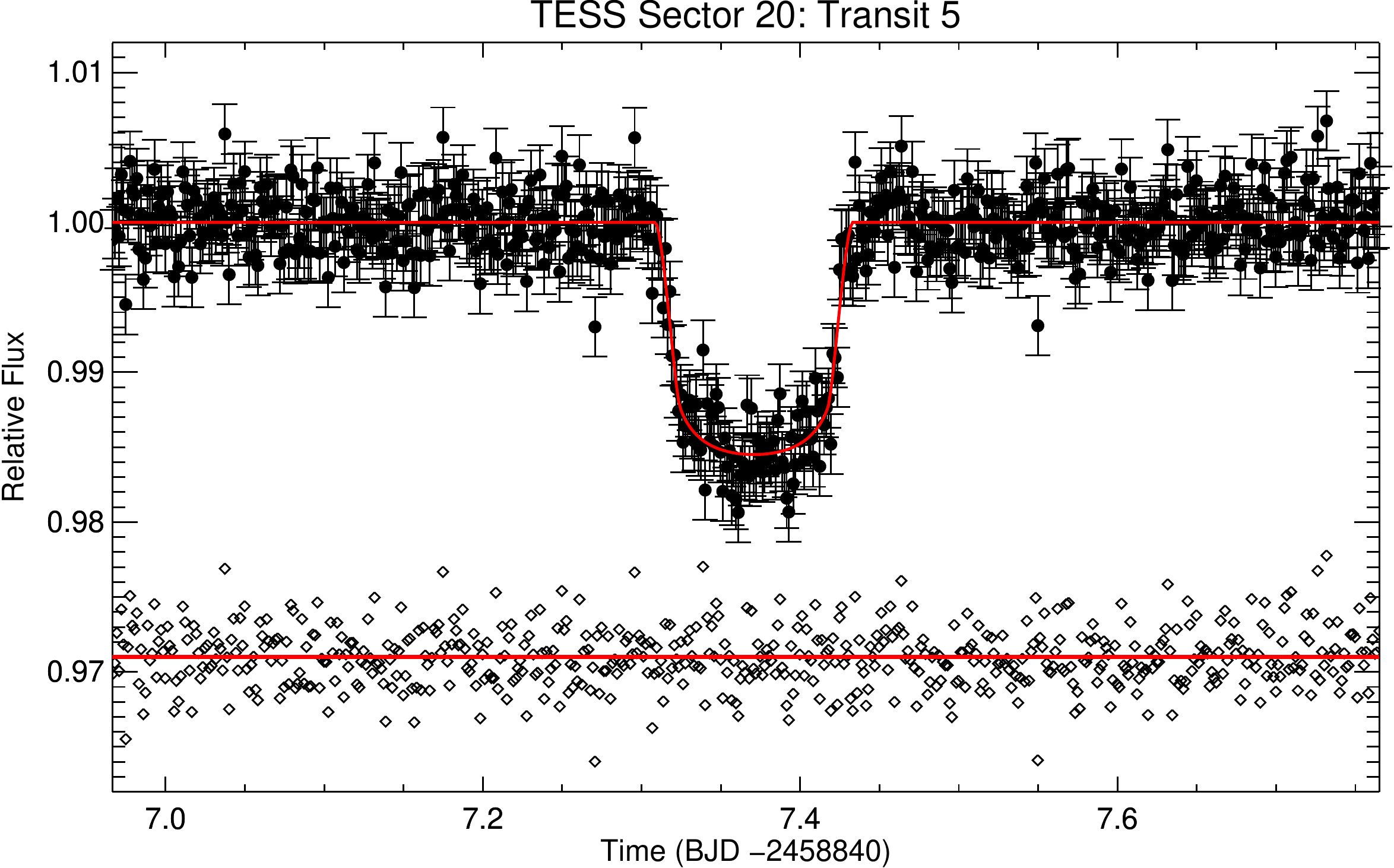} & \includegraphics[width=0.50\textwidth]{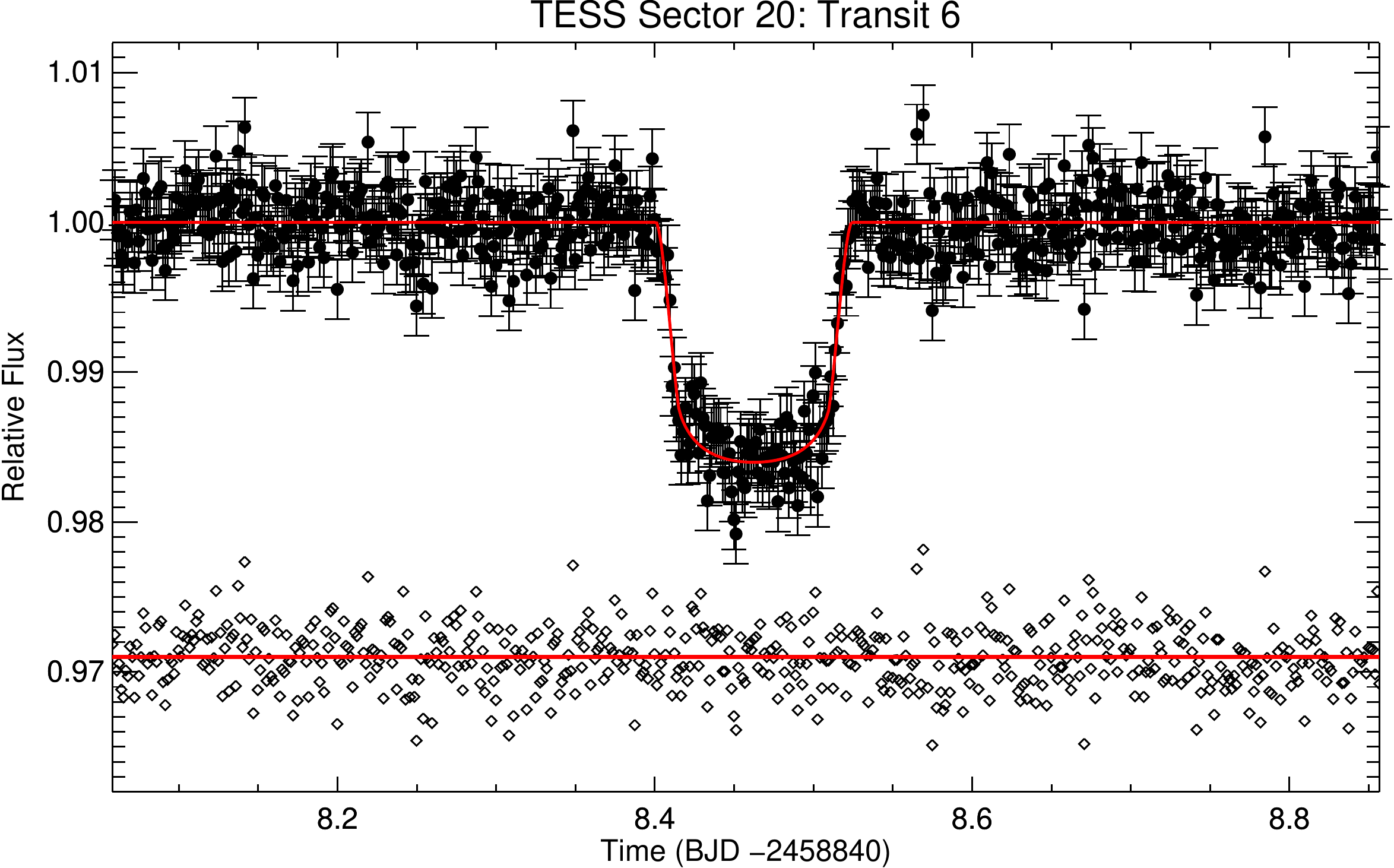}\\
  \end{tabular}
\caption{Individual TESS transit events (1-6) from Sector 20 of WASP-12b. The best-fitting model obtained from the EXOplanet MOdeling Package (\texttt{EXOMOP}) is shown as a solid red line. The residuals (light curve - model) are shown below the light curve.}
\label{fig:ind_transits_sec20_1}
\end{figure*}

\begin{figure*}
\centering
\begin{tabular}{cc}
    \includegraphics[width=0.50\textwidth]{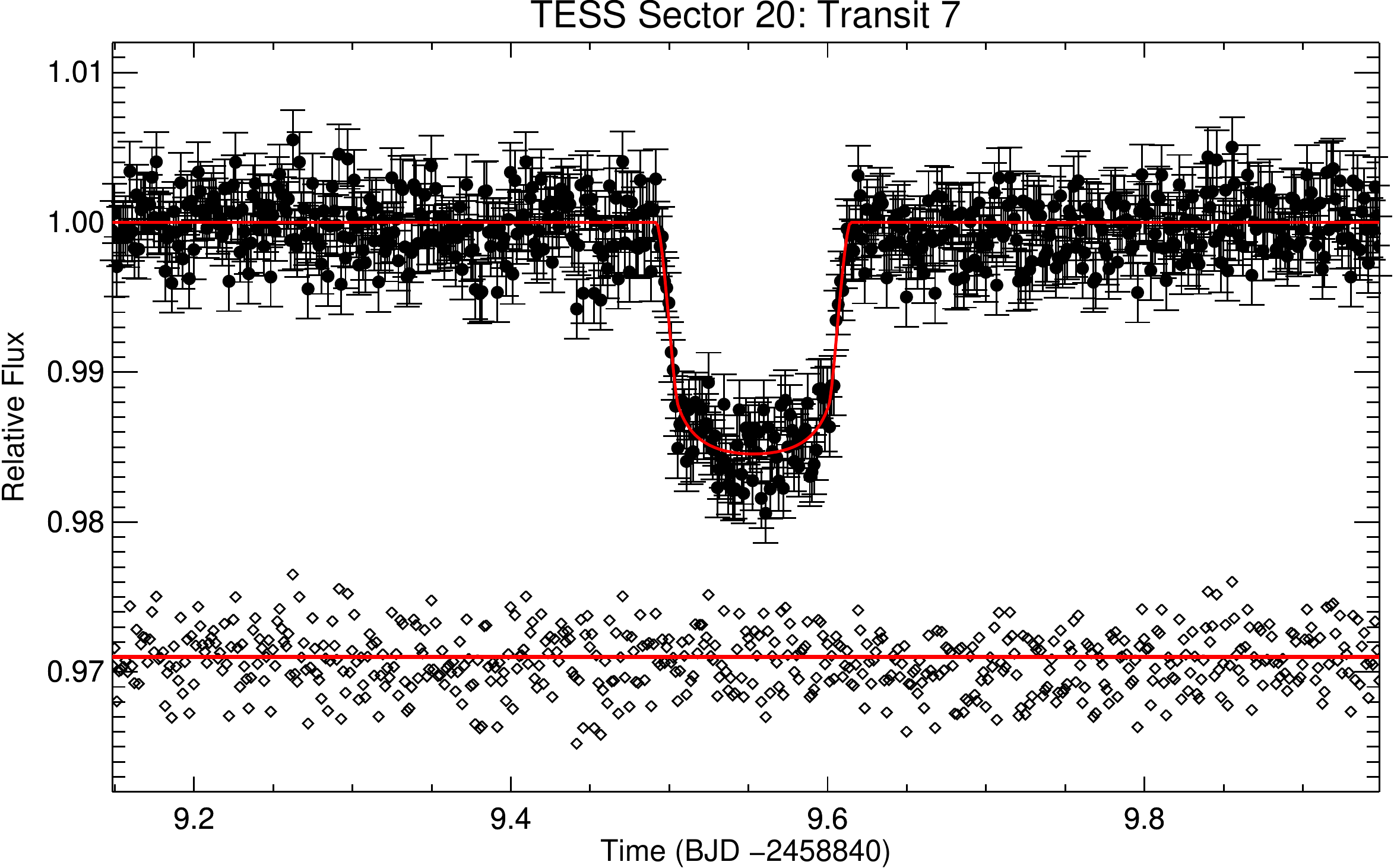} & \includegraphics[width=0.50\textwidth]{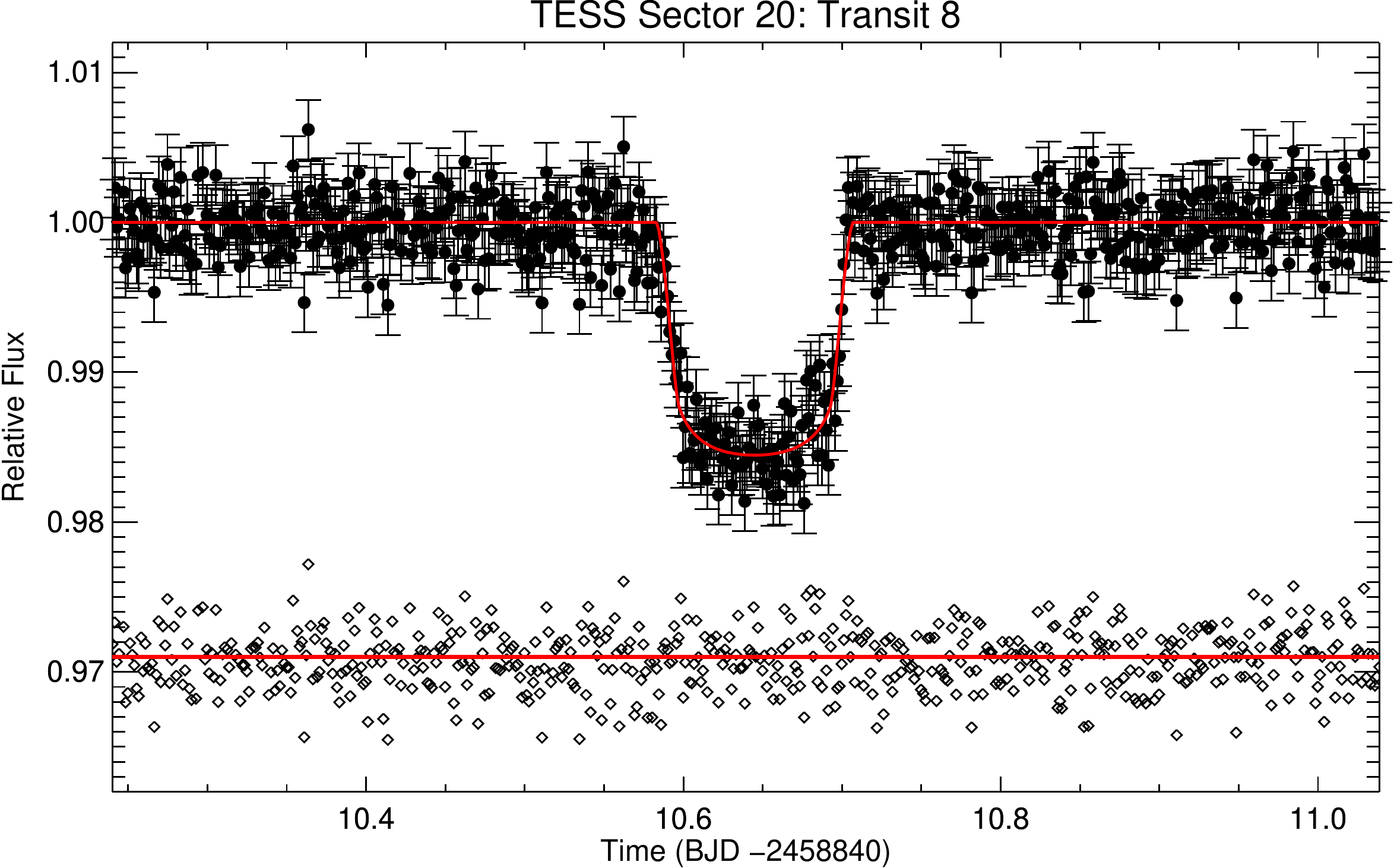}\\
 \includegraphics[width=0.50\textwidth]{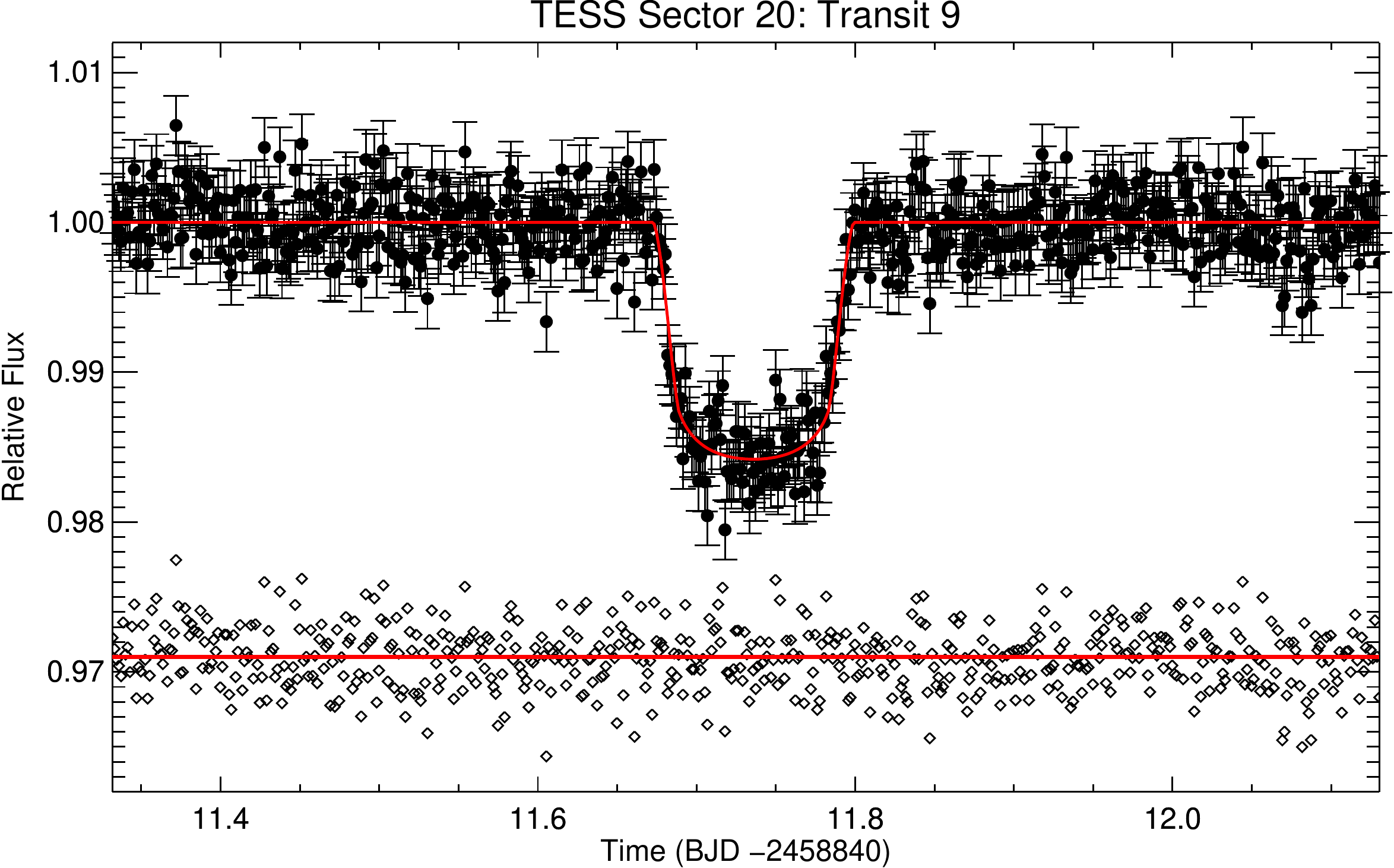} & \includegraphics[width=0.50\textwidth]{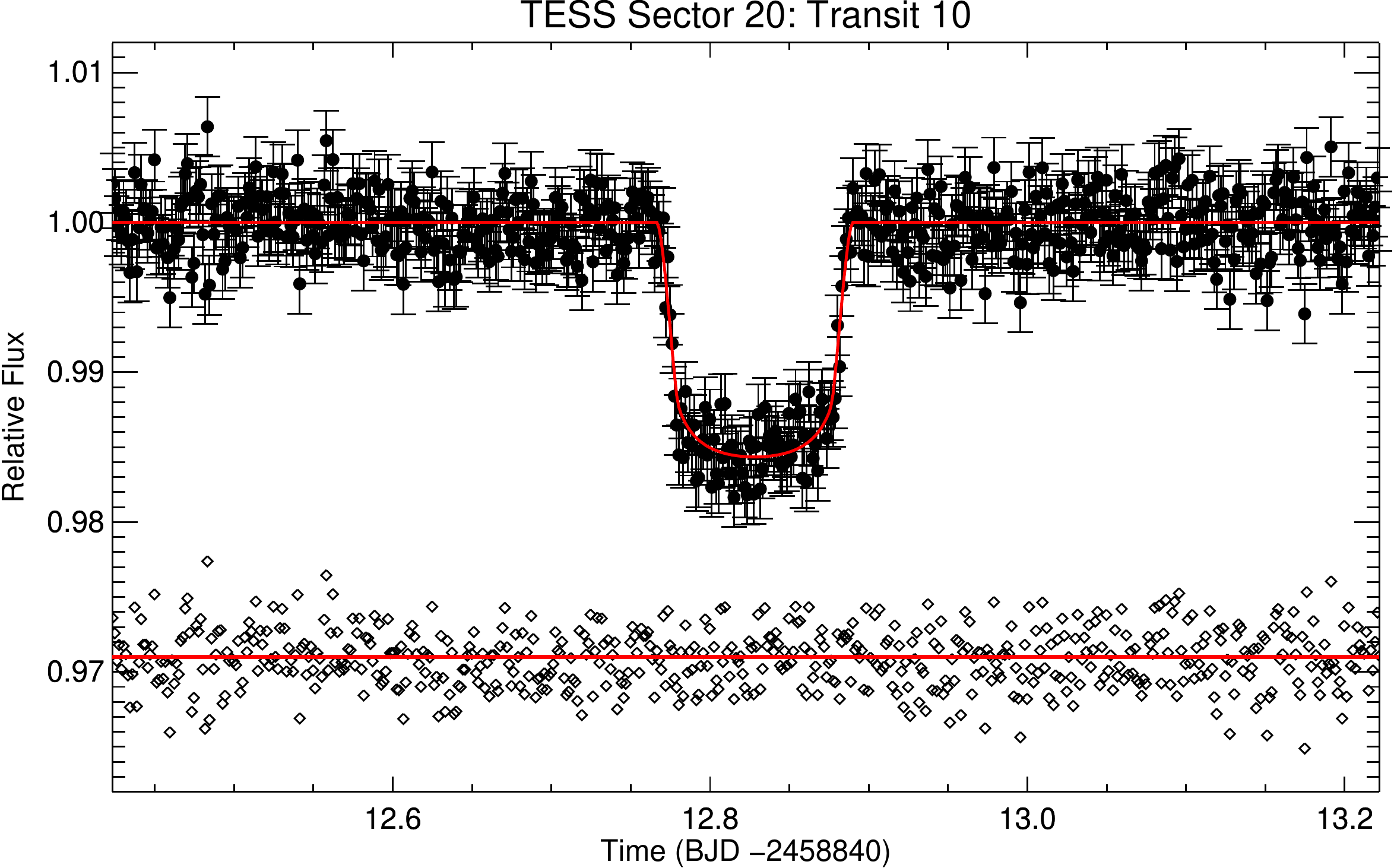}\\
 \includegraphics[width=0.50\textwidth]{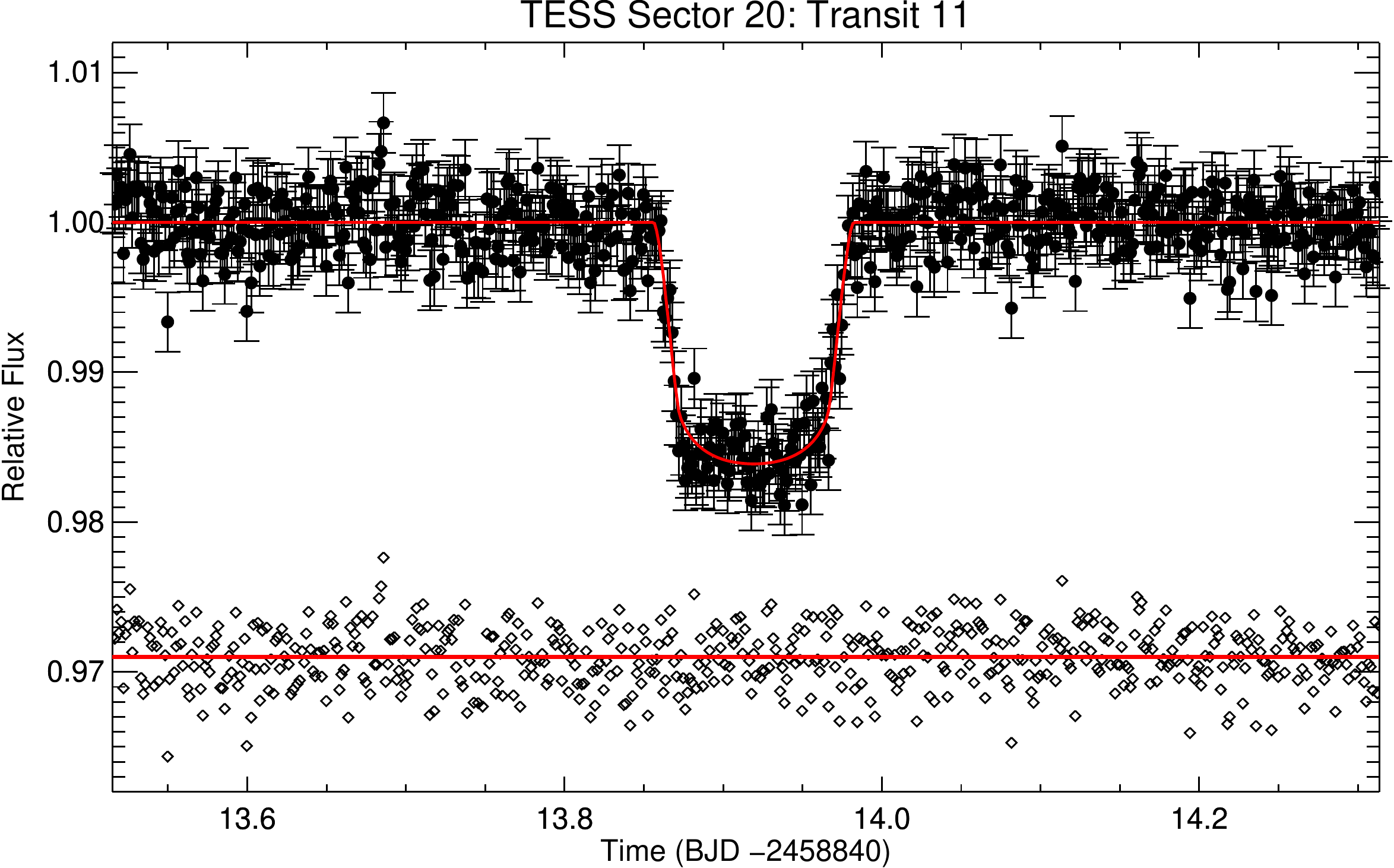} & \includegraphics[width=0.50\textwidth]{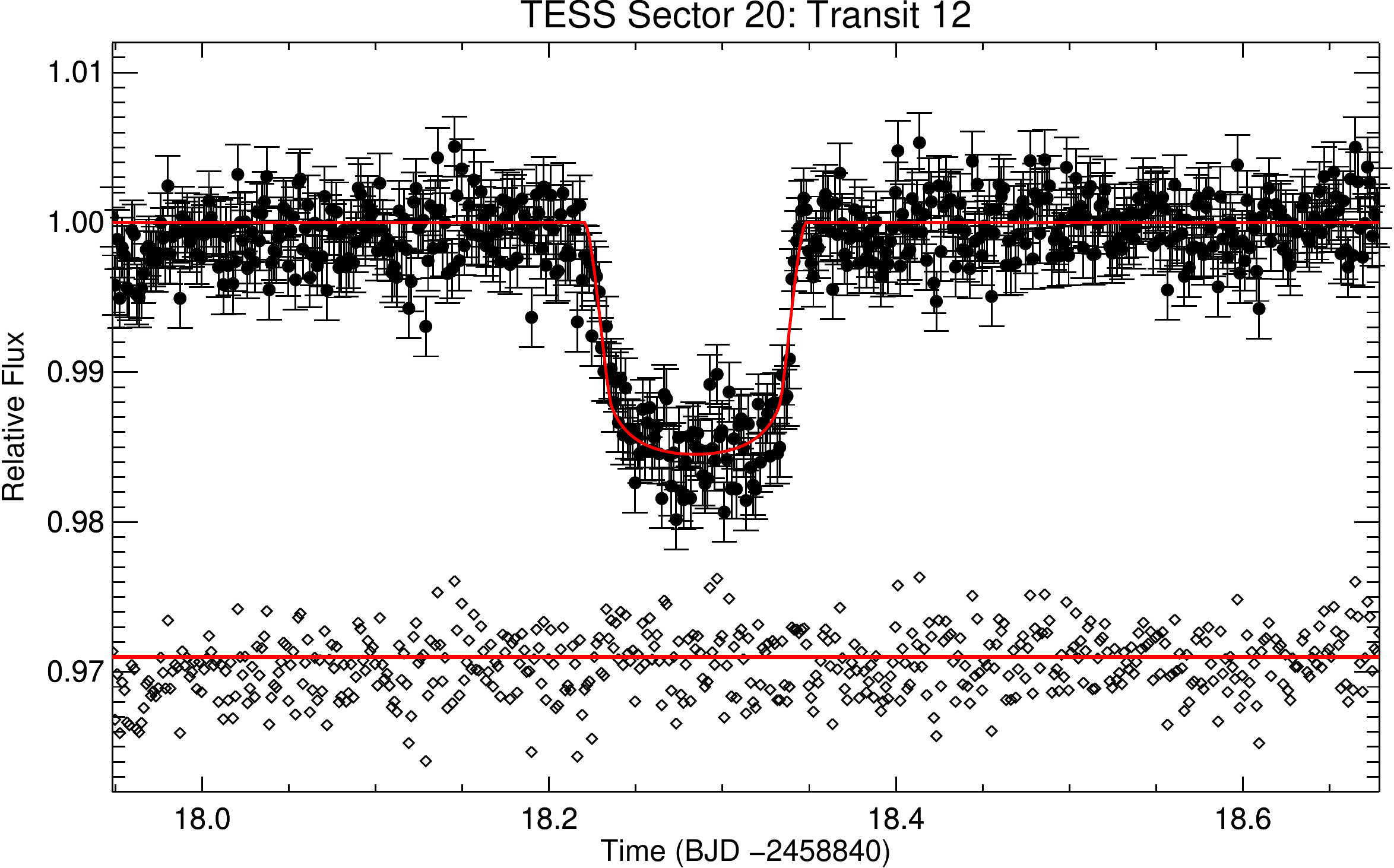}\\
  \end{tabular}
\caption{Individual TESS transit events (7-12) from Sector 20 of WASP-12b. Other comments are the same as Figure \ref{fig:ind_transits_sec20_1}.}
\label{fig:ind_transits_sec20_2}
\end{figure*}

\begin{figure*}
\centering
\begin{tabular}{cc}
  \includegraphics[width=0.50\textwidth]{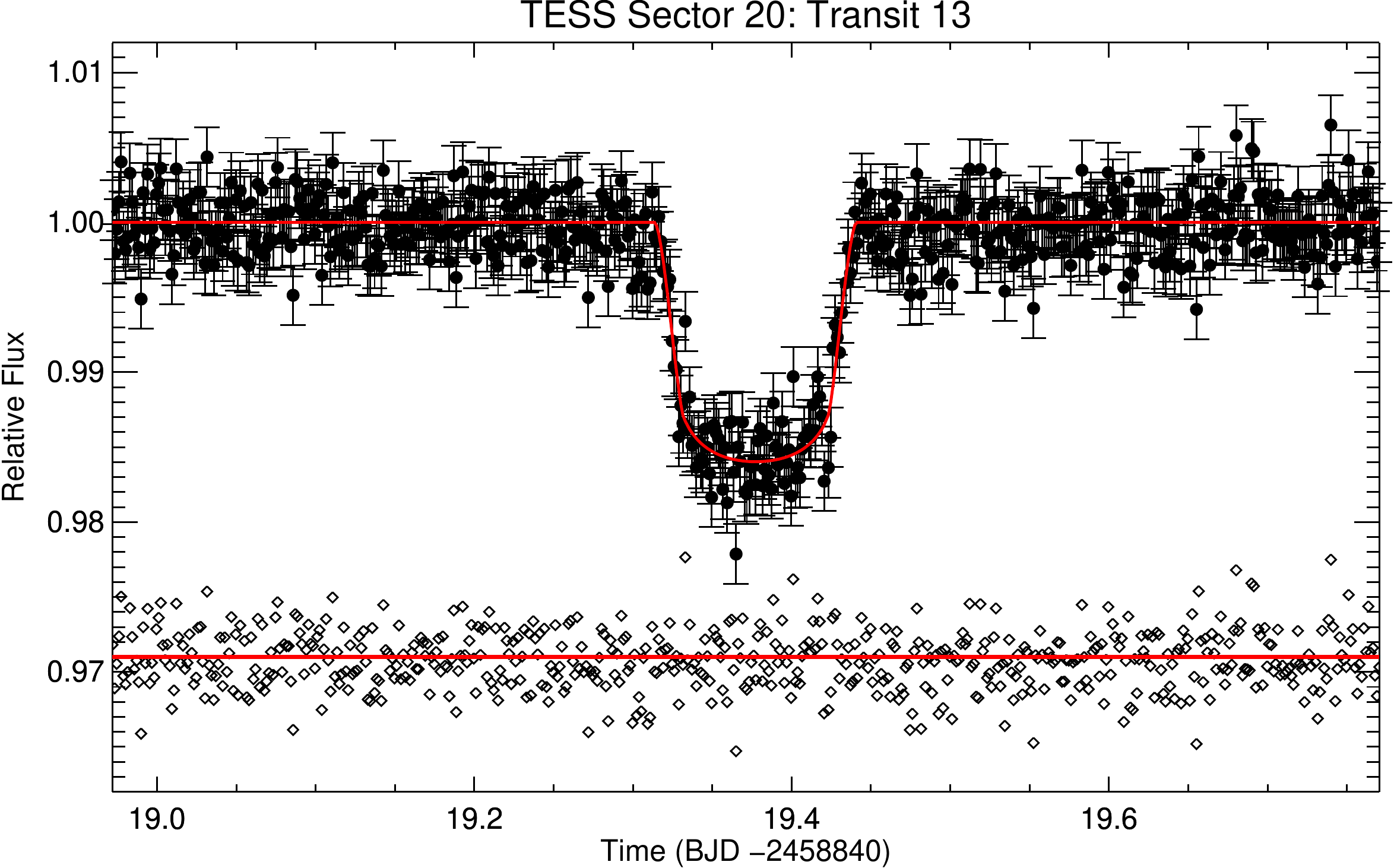} & \includegraphics[width=0.50\textwidth]{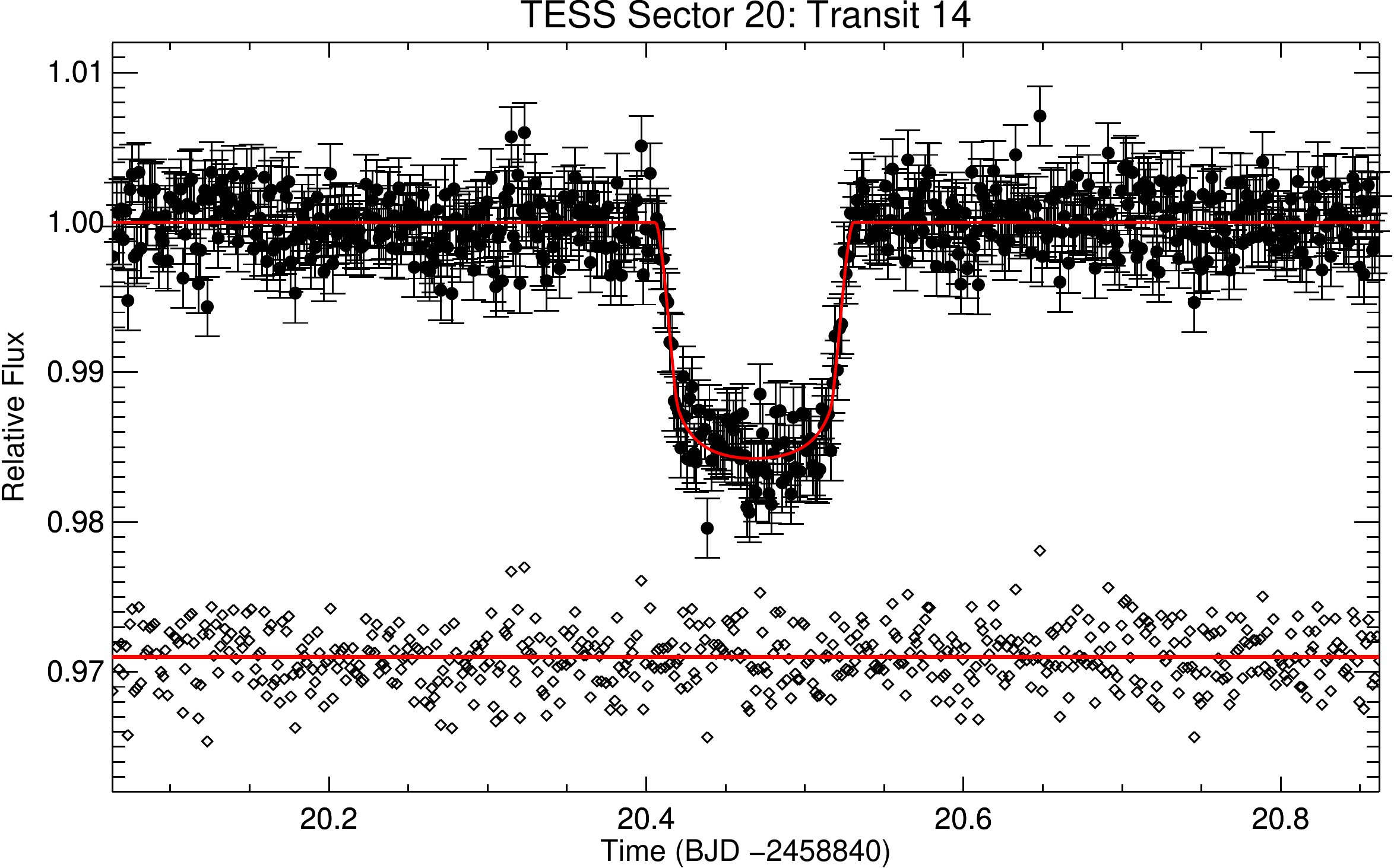}\\
    \includegraphics[width=0.50\textwidth]{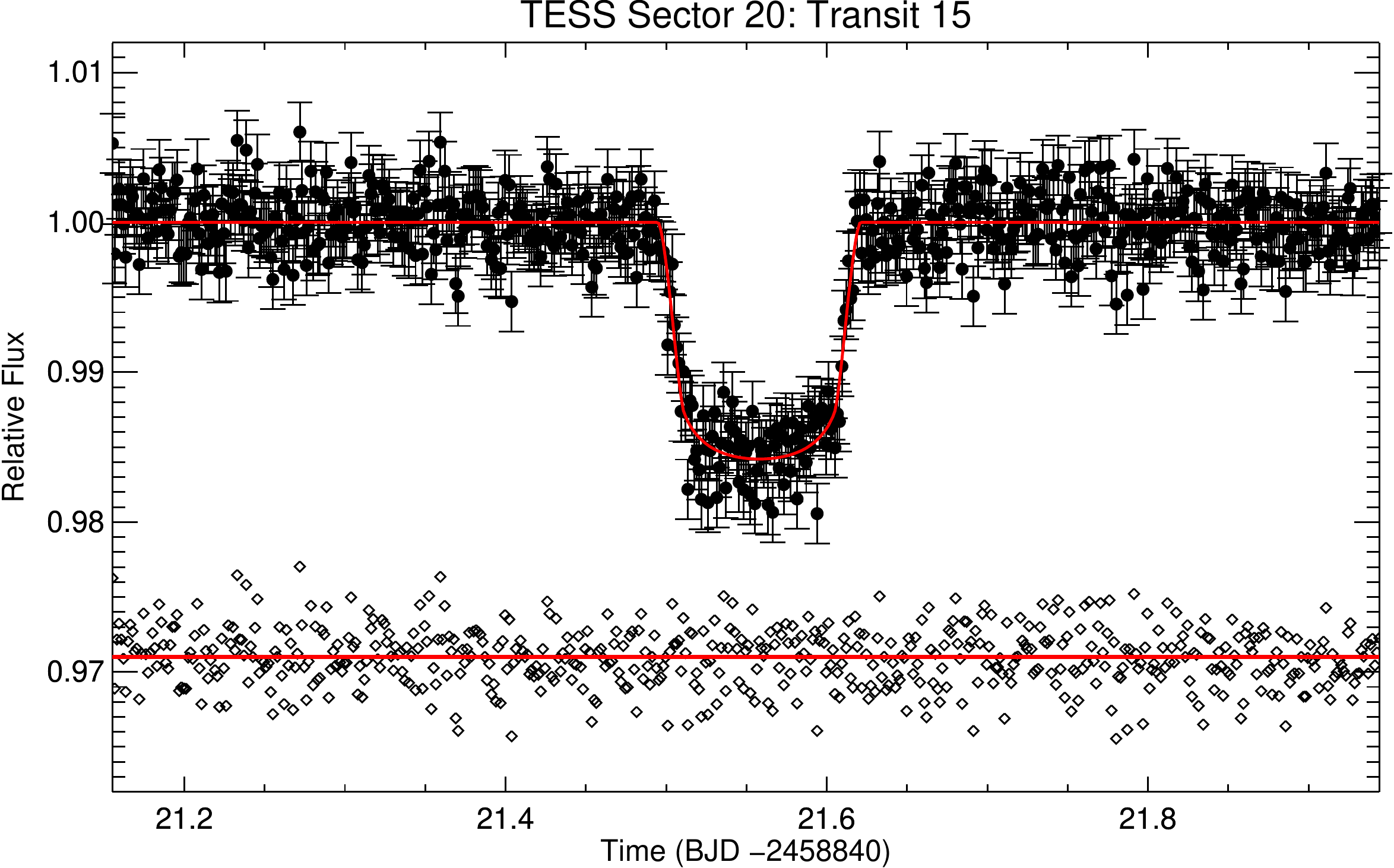} & \includegraphics[width=0.50\textwidth]{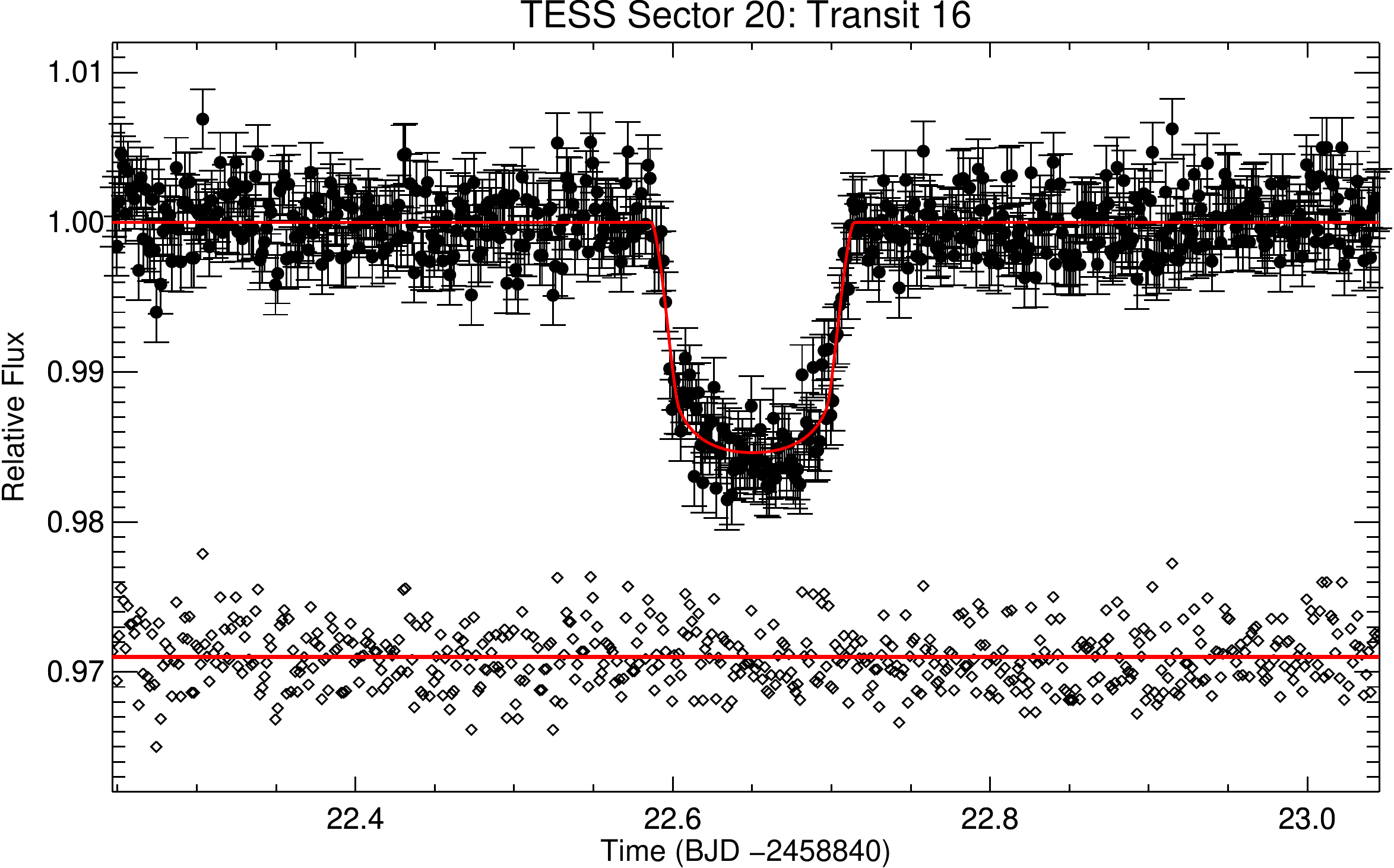}\\
 \includegraphics[width=0.50\textwidth]{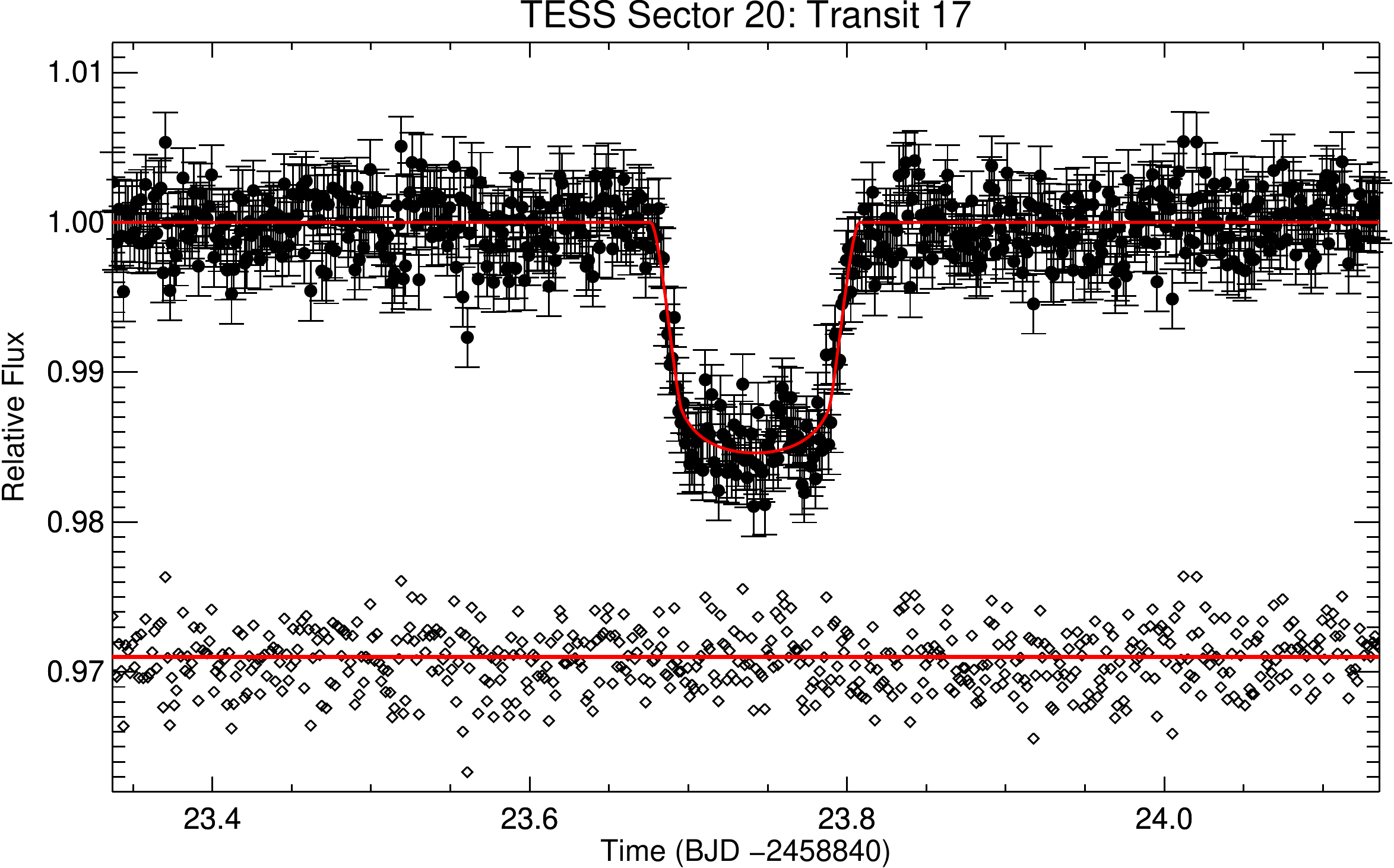} & \includegraphics[width=0.50\textwidth]{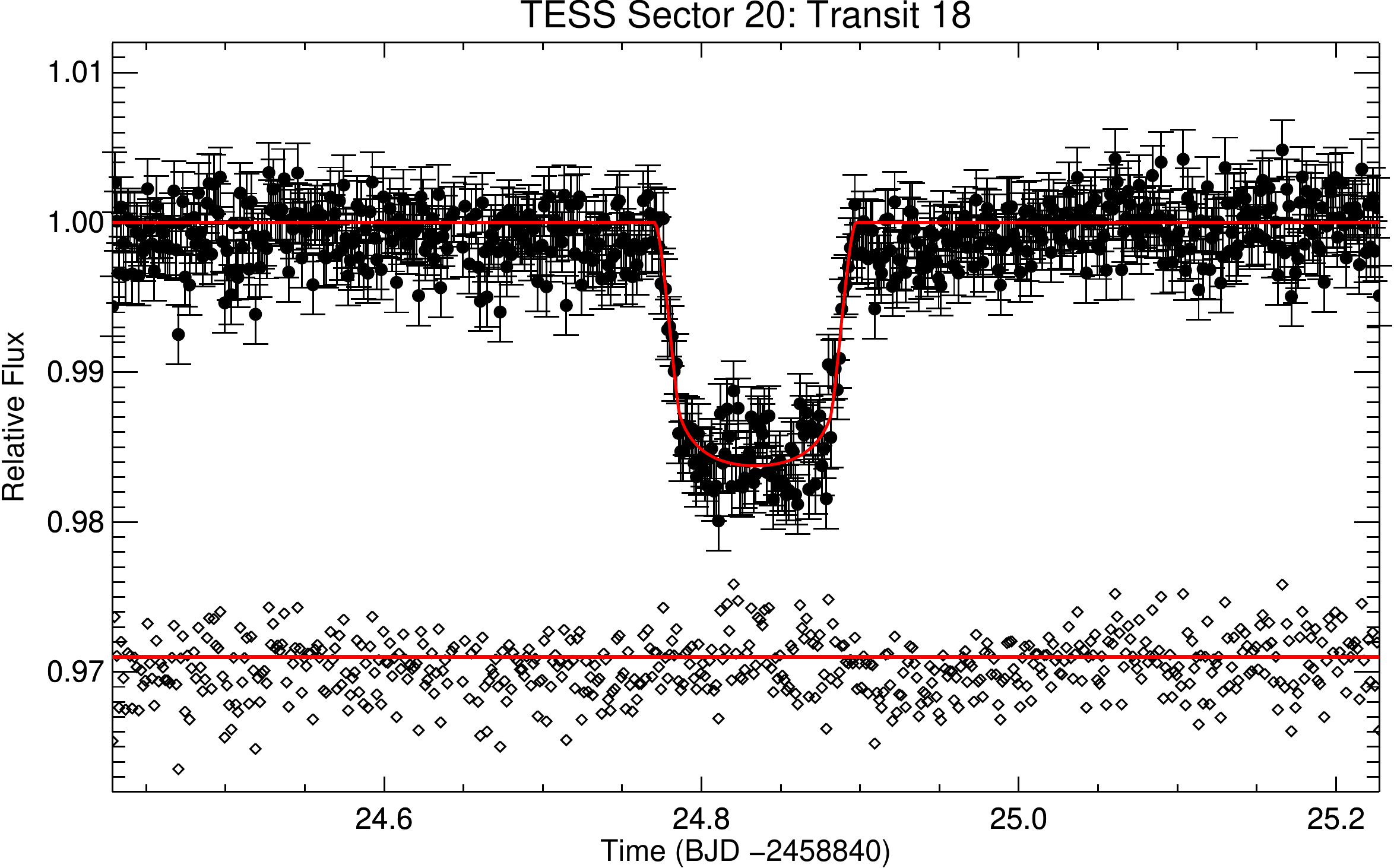}\\
  \end{tabular}
\caption{Individual TESS transit events (13-18) from Sector 20 of WASP-12b. Other comments are the same as Figure \ref{fig:ind_transits_sec20_1}.}
\label{fig:ind_transits_sec20_3}
\end{figure*}

\begin{figure*}
\centering
\begin{tabular}{cc}
\includegraphics[width=0.50\textwidth]{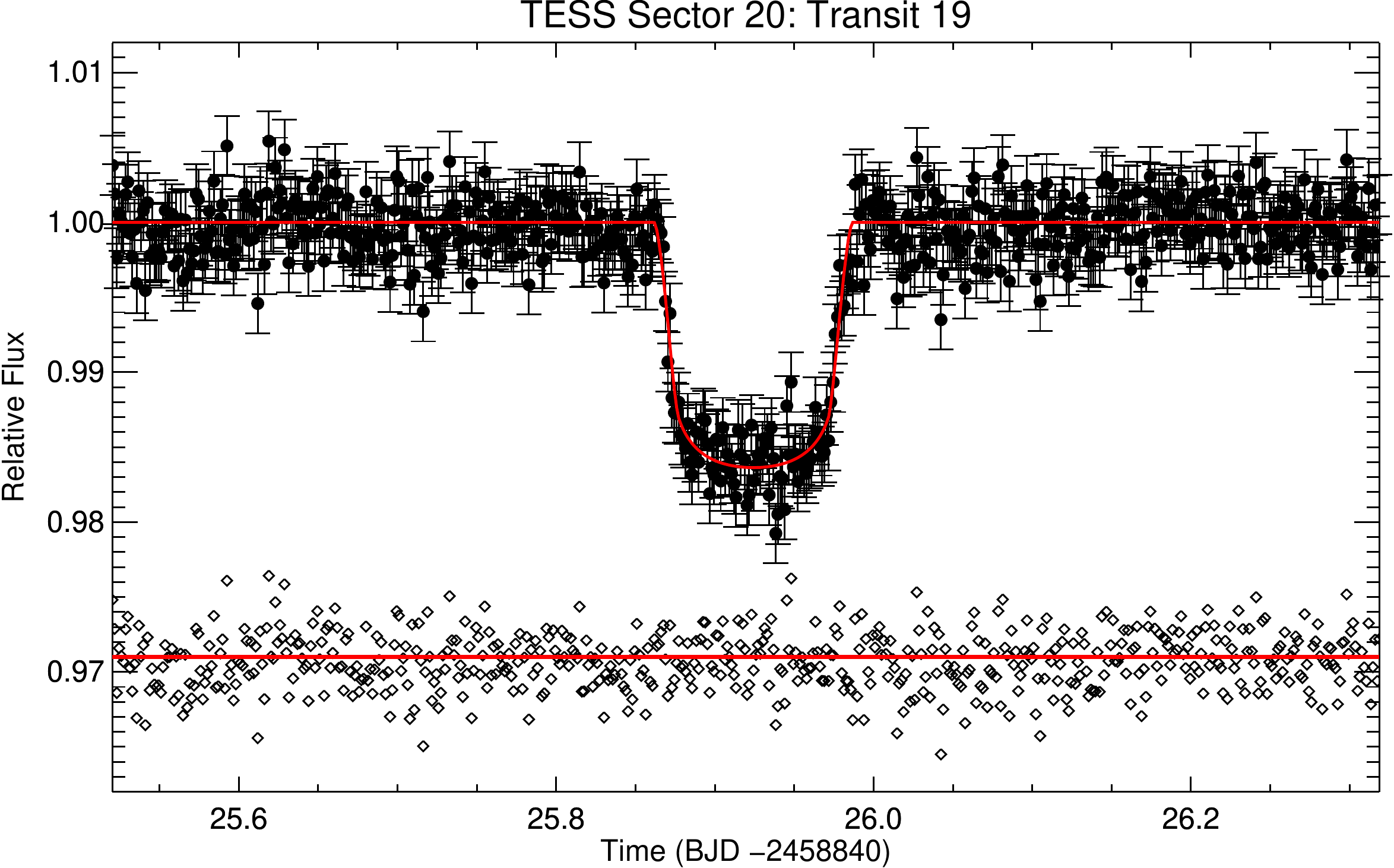} & \includegraphics[width=0.50\textwidth]{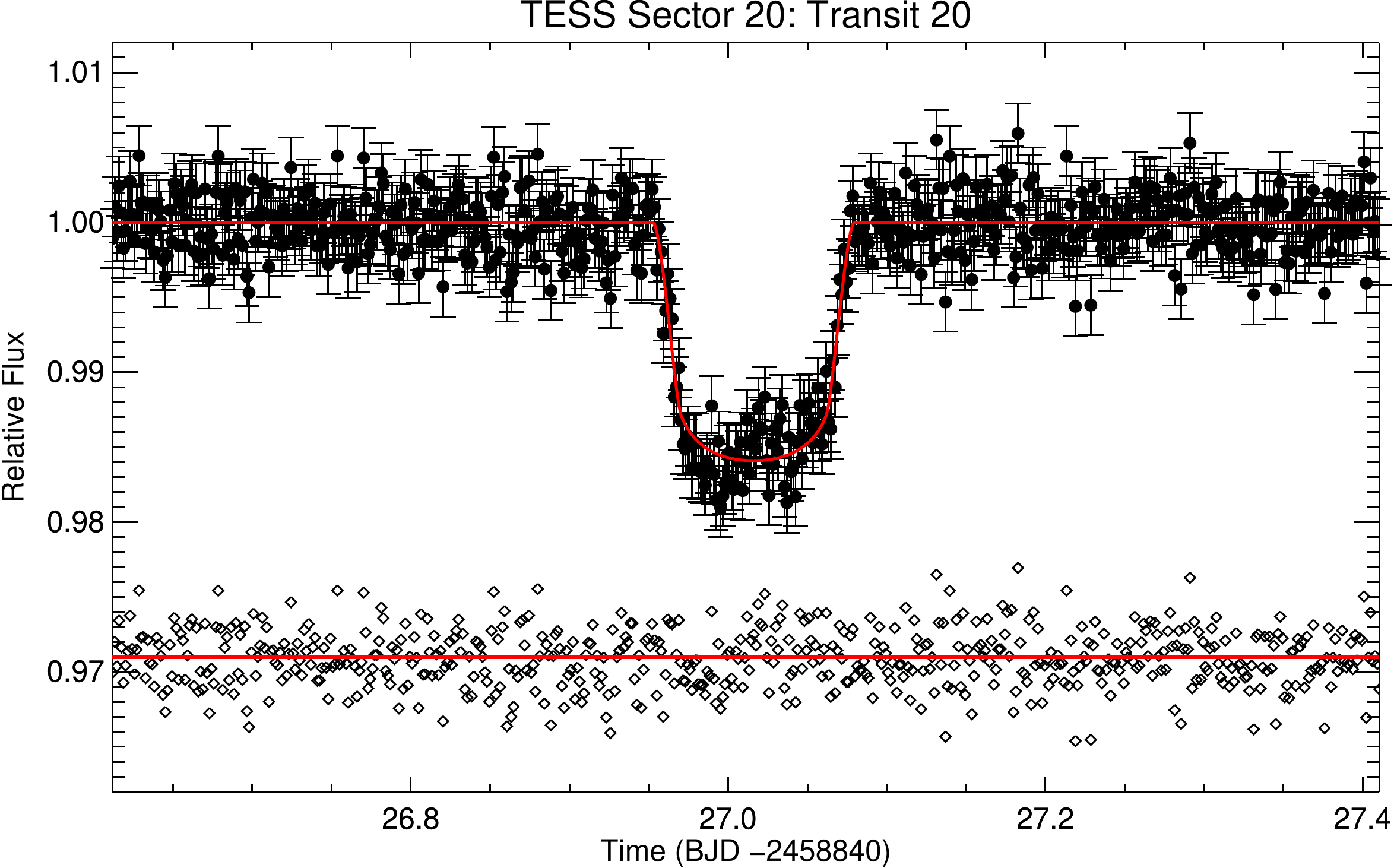}\\
   \end{tabular}
 \begin{tabular}{c}
  \includegraphics[width=0.50\textwidth]{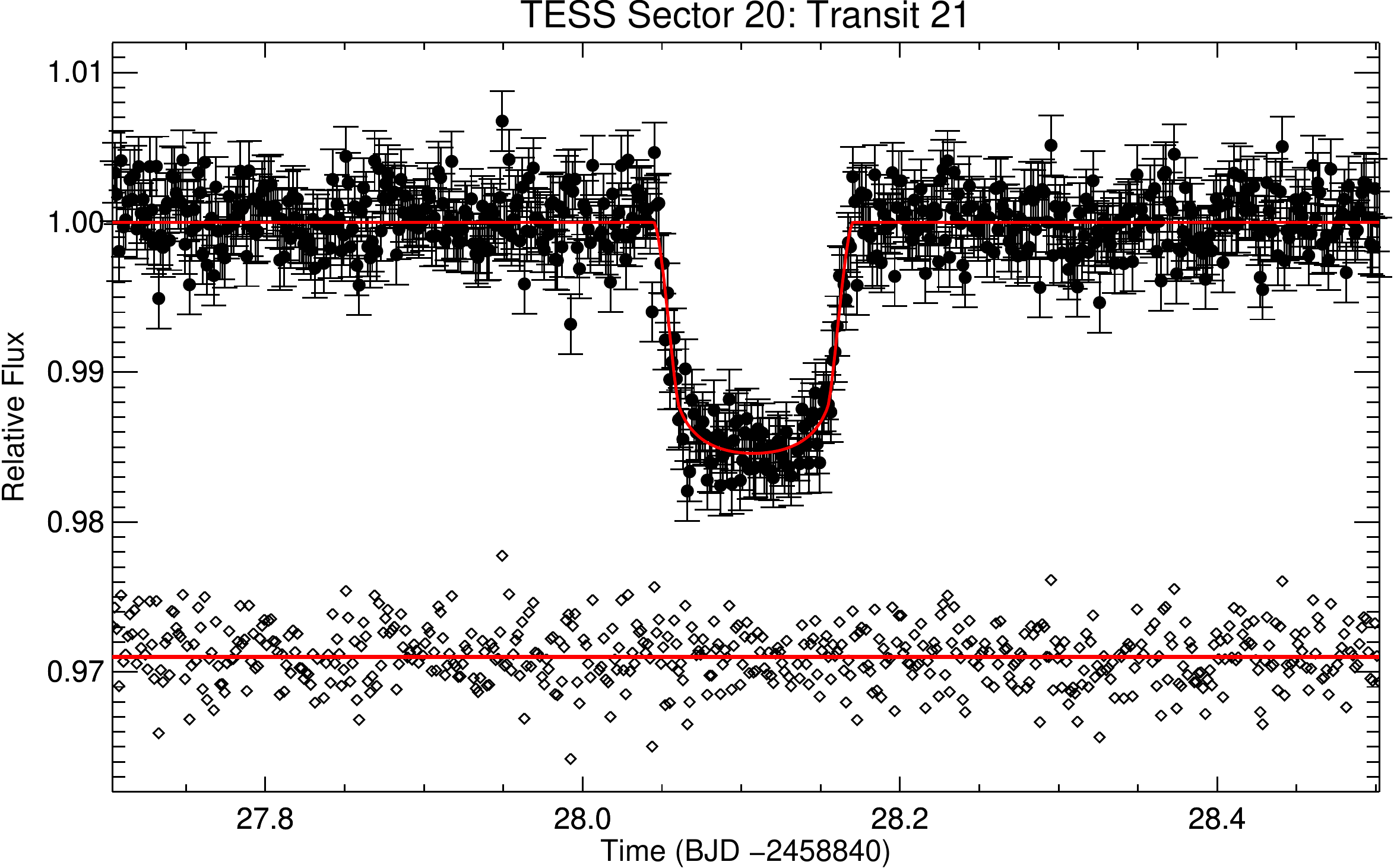} 
    \end{tabular}
\caption{Individual TESS transit events (19-21) from Sector 20 of WASP-12b. Other comments are the same as Figure \ref{fig:ind_transits_sec20_1}.}
\label{fig:ind_transits_sec20_4}
\end{figure*}

The parameters for each transit fit can be found in Table \ref{tb:lighcurve_model_TESS}. The light curves and \texttt{EXOMOP} model fits can be found in Figures \ref{fig:ind_transits_sec20_1}--\ref{fig:ind_transits_sec20_4}.

\begin{table}[htb]
\caption{Individual TESS Sector 20 transit parameters for WASP-12b derived using \texttt{EXOMOP}  }
    \centering
    \begin{tabular}{lccc}
    \hline
    \hline 
    Transit                         & 0                     & 1                     & 2     \\
    T$_{c}$ (BJD$_{TDB}$-2458840)   &  3.00484$\pm$0.00054  & 4.09663$\pm$0.00053   & 5.18779$\pm$0.00053                   \\
    R$_p$/R$_\ast$                  & 0.1229$\pm$0.0020     & 0.1207$\pm$0.0020     &0.1195$\pm$0.0019    \\
    a/R$_\ast$                      & 2.67$\pm$0.20         & 2.95$\pm$0.26         &2.92$\pm$0.22     \\
    Inclination (\degree)           & 77.25$\pm$3.08        & 81.26$\pm$2.72        &81.42$\pm$1.98        \\
    Duration (mins)                 & 185.98$\pm$3.00       & 177.89$\pm$2.87       & 180.00$\pm$2.83      \\
    \hline
    Transit                         & 3                       & 4                      & 5       \\
    T$_{c}$ (BJD$_{TDB}$-2458840)   &6.27961$\pm$0.00056      & 7.37087$\pm$0.00052   & 8.46245$\pm$0.00047    \\
    R$_p$/R$_\ast$                  &0.1220$\pm$0.0022        &0.1175$\pm$0.0018        & 0.1190$\pm$0.0013              \\
    a/R$_\ast$                      &2.76$\pm$0.16            & 3.05$\pm$0.23           & 3.14$\pm$0.11    \\
    Inclination (\degree)           &79.25$\pm$1.95           & 83.29$\pm$4.67          & 84.91$\pm$5.27       \\
    Duration (mins)                 & 185.98$\pm$2.83         & 175.96$\pm$2.83         &175.96$\pm$2.84       \\
    \hline
    Transit                         & 6                     & 7                     & 8         \\
    T$_{c}$ (BJD$_{TDB}$-2458840)   &9.55312$\pm$0.00043    &10.64494$\pm$0.00052     &11.73596$\pm$0.00055     \\
    R$_p$/R$_\ast$                  &0.1163$\pm$0.0010      &0.1169$\pm$0.0015      & 0.1188$\pm$0.0023              \\
    a/R$_\ast$                      &3.23$\pm$0.12          & 3.15$\pm$0.10           & 2.97$\pm$ 0.28                 \\
    Inclination (\degree)           &90.00$\pm$4.86         & 86.28$\pm$2.71          &82.45$\pm$6.12        \\
    Duration (mins)                 & 174.02$\pm$2.83       & 178.07$\pm$2.88         & 180.00$\pm$2.83      \\
    \hline
    Transit                         & 9                     & 10                    & 11        \\
    T$_{c}$ (BJD$_{TDB}$-2458840)   & 12.82820$\pm$0.00038  & 13.91930$\pm$0.00048    &18.28433$\pm$0.00064          \\
    R$_p$/R$_\ast$                  & 0.1178$\pm$0.0022      & 0.1201$\pm$0.0014       &0.1172$\pm$0.0016  \\
    a/R$_\ast$                      & 3.13$\pm$0.19           & 2.97$\pm$0.17           & 3.00$\pm$0.28                \\
    Inclination (\degree)           & 89.94$\pm$5.70          & 82.20$\pm$3.01          & 83.91$\pm$4.60       \\
    Duration (mins)                 & 174.02$\pm$2.83       & 180.00$\pm$2.85         & 180.00$\pm$2.85      \\
     \hline
    Transit                         & 12                    & 13                        & 14      \\
    T$_{c}$ (BJD$_{TDB}$-2458840)   & 19.37734$\pm$0.00054    & 20.46819$\pm$0.00043    & 21.55796$\pm$0.00049    \\
    R$_p$/R$_\ast$                  & 0.1202$\pm$0.0017       & 0.1176$\pm$0.0011       &0.1186$\pm$ 0.0015               \\
    a/R$_\ast$                      & 2.87$\pm$0.19           & 3.17$\pm$0.15           &3.01$\pm$0.20   \\
    Inclination (\degree)           & 80.40$\pm$2.34          & 87.55$\pm$2.92          & 83.09$\pm$4.09       \\
    Duration (mins)                 & 180.00$\pm$2.85         & 177.89$\pm$2.85         & 180.00$\pm$2.83       \\
    \hline
    Transit                         & 15                    & 16                        & 17      \\
    T$_{c}$ (BJD$_{TDB}$-2458840)   & 22.65043$\pm$0.00058  &23.74207$\pm$0.00059       &24.83394$\pm$0.00048     \\
    R$_p$/R$_\ast$                  & 0.1175$\pm$0.0019     & 0.1185$\pm$0.0020         &0.1202$\pm$0.0015               \\
    a/R$_\ast$                      & 2.90$\pm$0.30         & 2.73$\pm$0.24             &3.00$\pm$0.17          \\
    Inclination (\degree)           & 81.30$\pm$6.06        & 78.80$\pm$3.43            &83.33$\pm$5.16       \\
    Duration (mins)                 & 182.11$\pm$2.87       & 186.00$\pm$2.83           &180.00$\pm$2.86       \\
    \hline
     Transit                         & 18                   & 19                        & 20                  \\
    T$_{c}$ (BJD$_{TDB}$-2458840)   & 25.92430$\pm$0.00051  & 27.01610$\pm$0.000 49     & 28.10764$\pm$0.00046    \\
    R$_p$/R$_\ast$                  & 0.1203$\pm$0.0014     & 0.1197$\pm$0.0018         & 0.1173$\pm$0.0014 \\
    a/R$_\ast$                      & 3.09$\pm$0.12         & 2.92$\pm$0.19             & 3.01$\pm$0.14                \\
    Inclination (\degree)           &84.77$\pm$12.04        &81.12$\pm$3.02             &82.85$\pm$2.77        \\
    Duration (mins)                 &178.07$\pm$2.83        &180.00$\pm$2.83            & 180.00$\pm$2.90      \\
    \hline
    \hline
    \end{tabular}
    \tablecomments{The linear and quadratic limb darkening coefficient used in the analysis are 0.2131 and 0.3212 \citep{Claret2017}}
    \label{tb:lighcurve_model_TESS}
\end{table}

\newpage
\clearpage 
\bibliographystyle{aasjournal} 
\bibliography{reference.bib} 



\end{document}